\documentclass[twocolumn]{aastex62}
\usepackage{float,graphicx,amsmath,multirow,mathtools}
\usepackage{color}
\usepackage[version=4]{mhchem}
\usepackage{comment}
\usepackage{amsmath}
\usepackage[caption=false]{subfig}
\usepackage{subfig}

\usepackage{bm}

\newcommand{\kms}{km~s$^{-1}$}

\newcommand{\msunyr}{M$_{\Sun}$\ yr$^{-1}$}

\newcommand{\cmnegtwo}{cm$^{-2}$}
\newcommand{\cmnegthree}{cm$^{-3}$}

\DeclareCaptionFormat{cont}{#1 (cont.)#2#3\par}

\begin{document}

\title{Molecular Distributions and Abundances in the Binary-Shaped Outflow of V Hya}
\author{Mark A. Siebert}
\affiliation{Department of Astronomy, University of Virginia, Charlottesville, VA 22904, USA}
\affiliation{Department of Space, Earth and Environment, Chalmers University of Technology, SE-412 96 Gothenburg, Sweden}
\author{Raghvendra Sahai}
\affiliation{Jet Propulsion Laboratory, California Institute of Technology, MS 183-900, Pasadena, CA 91109, USA}
\author{Samantha Scibelli}
\altaffiliation{Jansky Fellow of the National Radio Astronomy Observatory.}
\author{Anthony J. Remijan}
\affiliation{National Radio Astronomy Observatory, Charlottesville, VA 22903, USA}

\begin{abstract}
Binaries are known to play a key role in the mass loss and dynamical environments of evolved stars. Stellar and sub-stellar companion interactions produce complex wind morphologies including rotating/expanding disks, bipolar outflows, and spiral wind patterns; however, the connection between these many structures and the gas phase chemistry they harbor is not well-constrained. To expand the sample of chemical inventories in interacting systems, we present a detailed spectroscopic case study of the binary C-rich Asymptotic Giant Branch (AGB) star V Hya. Using spatially resolved ALMA observations at Bands 3, 6 and 7, we characterize the rotational emission lines and distributions of molecules in its surrounding disk undergoing dynamical expansion (DUDE). We detect emission from over 15 molecules and isotopologues toward this source, and present resolved maps for the brightest tracers of carbonaceous chemistry (e.g. \ce{CCH}, \ce{C4H}, \ce{HC5N}, \ce{HNC}, \ce{CH3CN}). Employing LTE and non-LTE models of emission from the DUDE, we estimate the abundance distributions for optically thin species, and compare them with prototypical carbon-rich AGB envelopes. We find that the average abundances of detected species are within a factor of ${\sim}5$ from sources with similar mass-loss rates; however, the distribution of daughter species in V Hya is much more compact, with carbon chain species (\ce{CCH}, \ce{C4H}, \ce{HC3N}) appearing with abundances $>$10$^{-7}$ even in the innermost sampled regions (200\,au) of the disk. 

\end{abstract}

\submitjournal{ApJ}
\accepted{November 1, 2024}

\section{Introduction}
The circumstellar envelopes of Asymptotic Giant Branch (AGB) stars are known for their rich molecular composition and their instrumental role in the chemical enrichment of the interstellar medium (ISM). It is estimated that these evolved stars contribute ${\sim}$80\% of recycled interstellar gas, ${\sim}$30\% of newly formed dust grains, in addition to a significant fraction of aromatic molecules \citep{Tielens2005_DustProductionISM,Zeichner2023_RyuguPAHs}. The type of enrichment is primarily dependent on the surface chemistry of the progenitor star, as circumstellar outflows typically exhibit either oxygen-rich (C/O$<$1) or carbon-rich (C/O$>$1) chemistry. In the traditional picture of circumstellar chemistry, gas phase molecules in AGB envelopes are produced in two stages, with parent species forming in the extended stellar atmosphere of the star. Then as they travel outward and the density decreases, kinetics and photochemistry dominate, producing a wide variety of daughter molecules \citep[e.g.\ ][]{Hofner2018_AGBreview}. 

For C-rich environments, circumstellar chemistry is primarily understood through observations of the nearby carbon star IRC+10216 and its rich chemistry comprising over 100 uniquely identified molecules \citep{McGuire2022_census}; however, the applicability of this example to all other sources is still largely unconstrained. \citet{Unnikrishnan2024_charting} recently studied the chemical inventories of three carbon stars with comparable mass-loss rates to IRC+10216 (II Lup, V358 Lup, and AI Vol), finding molecular abundances consistent across sources, indicating that IRC+10216 is indeed a good prototype for this chemistry. A remaining question; however, is whether this molecular framework changes when moving to more dynamically complex stellar wind structures that we now know to be commonplace.

Recently, the prevalence of binary interactions has greatly impacted our understanding of dynamical evolution in AGB and Red Supergiant (RSG) environments. The ATOMIUM\footnote{ALMA tracing the origins of molecules forming dust in oxygen-rich M-type stars} survey revealed a vast array of AGB wind morphologies likely induced by stellar and substellar companions \citep{Decin2020_atomium}. These structures can be broadly categorized into two groups, depending on orbital characteristics of the system. Spirals structures are associated with high mass-loss rates and wide companions, and can be observed in objects such as CIT 6, AFGL 3068, and R Scl \citep{Kim2015_CIT6Bipolar,Guerrero2020_ringsarcs_pne,Maercker2016_RScl_shell}. In contrast, equatorial density enhancements and disks are more prominent in short period systems, like $\pi^1$ Gru and R Aqr \citep{Chiu2006_piGru,Liimets2018_RAqr_morph}. 

In addition to the dynamical shaping, stellar companions are believed to have an impact on the chemistry of their surrounding envelope. This was first found theoretically by \citet{VandeSande2022}, who showed that embedded binaries can provide sufficient visible and UV flux to drive photochemistry at locations in the envelope where it is usually inefficient. \citet{Siebert2022_IRC_HC3N} later invoked a solar-type binary companion to reproduce observed abundances of \ce{HC3N} in the inner layers of IRC+10216. Furthermore, \citet{Danilovich2024_WAql_binarychem} recently found evidence of companion-driven chemistry imprinted in the outflow of the highly elliptical system W Aql. The full scope of binary-outflow chemical interactions are still not well-constrained, and given the range of physical structures produced in these environments, unbiased line surveys of objects at different stages of evolution are crucial to building up this picture.

V Hya is a C-rich AGB star at a distance of 400\,pc \citep{Sahai2022_VHyaDUDE} with a strongly-interacting, short-period stellar companion \citep[$P{\sim}$17\,yr, $a{\sim}$11\,au][]{Planquart2024_VHya_comp_monitoring}. It displays a far-UV excess which could be associated with this companion or an accompanying accretion disk \citep{Sahai2008_GALEX_UV_AGB}. On much larger scales, V Hya is surrounded by a disk undergoing dynamical expansion (DUDE) comprising several concentric toroidal rings, in addition to episodic bipolar bullet-like ejections \citep{Hirano2004_VHya,Sahai2009_VHya,Sahai2016_VHyaHST,Scibelli2019_VHya_bulletModeling}. The observed total mass-loss rate of these components is quite high \citep[$2.45\times10^{-5}$\,\msunyr; ][]{Knapp1997_VHya}\footnote{Sum of mass-loss rates derived from three observed velocity components at 15, 45, and 200\,\kms}, indicating it is a very evolved source nearing the end of the AGB. The equatorial structure and hints of bipolarity in V Hya suggest the beginnings of the much more dramatic morphologies that are common in proto-planetary nebulae (PPNe) \citep{Olofsson2019_HD101584, Sahai2017_IRAS05506_bullets}. The dense waists of these post-AGB objects have been shown as efficient sites of dust growth during this period of rapid stellar evolution \citep{Sahai2006_I22036_graintorus}, so characterizing molecular processing in their nascent stages is relevant to understanding their overall chemical yields.

V Hya is a unique transitory source that is well-poised to reveal new insights into chemistry in this important type of binary-induced geometry. In this paper, we present an interferometric spectral line study of this source using available ALMA data to constrain its gas phase molecular inventory. Using these observations we aim to characterize the dominant chemical processes throughout this unique source, and compare it with previous studies of more spherical C-rich AGB targets.

In Section \ref{section:observations}, we summarize the observations and data reduction. In Section \ref{section:results}, we present the spectra and detected molecular lines, as well as spatially resolved maps of key species. Finally, in Section \ref{section:disc}, we discuss excitation mechanisms, and use multiple methods to derive abundances of chemical species in the disk of V Hya.

\begin{deluxetable*}{ccccccccc}[t!]
    \tablecaption{Summary and characteristics of the ALMA data utilized in this work. }
    \tablewidth{\columnwidth}
    \tablehead{
    \colhead{Project code} & \colhead{Band} &\colhead{Frequency range} &\colhead{Obs.\ date(s)} & \colhead{Array(s)}  &\colhead{Synthesized beam} & \colhead{Spectral res.} &\colhead{Image RMS}\\  & & (GHz) & & &($\theta_{\mathrm{maj}}\times\theta_{\mathrm{min}}$) & (\kms) & (mJy beam$^{-1}$)} 
    \startdata
	2015.1.01271.S & 3 & 87.2 -- 91.0 &Mar. -- Aug.\ 2016 & 12m$^a$ & $1.1\arcsec\times0.90\arcsec$  & 3.2  & 1.2 \\
	  & 3 & 99.1 -- 102.9 &Mar. -- Aug.\ 2016 & 12m$^a$ & $0.89\arcsec\times0.74\arcsec$  & 2.9  & 1.1 \\
	2018.1.01113.S & 6 & 218.6 -- 220.5 & Apr.\ 2020 & 12m\phantom{$^a$} & $0.64\arcsec\times0.54\arcsec$ & 0.66  & 1.1 \\
	 & 6 & 220.2 -- 220.6 & Apr.\ 2020 & 12m\phantom{$^a$} & $0.64\arcsec\times0.54\arcsec$ & 0.33  & 1.8 \\
         & 7 &  330.2 -- 332.2 & Mar.\ -- Apr.\ 2020 & 12m+7m & $0.51\arcsec\times0.40\arcsec$  & 0.44  & 1.5 \\
         & 7 &  342.3 -- 344.2 & Mar.\ -- Apr.\ 2020 & 12m+7m & $0.50\arcsec\times0.39\arcsec$  & 0.43  & 1.1 \\
         & 7 &  345.55 -- 346.02 & Mar.\ -- Apr.\ 2020 & 12m+7m & $0.49\arcsec\times0.39\arcsec$  & 0.10  & 2.8 
    \enddata   
    \tablecomments{$^a$Includes two data sets from the 12m array in short and long baseline configurations}
    \label{table:obspar-ALMA}
\end{deluxetable*}

\section{Observations and reduction}
\label{section:observations}
For this investigation, we utilize data from two archival programs from the Atacama Large Millimeter/submillimeter Array (ALMA). A summary of the frequency coverage, as well as the obtained spectral sensitivity and resolution is shown Table \ref{table:obspar-ALMA}. Project 2015.1.01271.S (PI: Keller) includes observations of seven carbon stars using the Band 3 receiver (84 -- 116\,GHz). Two of the sources from this data set were analyzed by \citet{Unnikrishnan2024_charting}, who detected and mapped the distributions carbon chemistry products (e.g.\ \ce{HC3N}, \ce{HC5N}, \ce{CN}). The observations of V Hya in this data set comprise executions using compact (80th percentile baseline length -- 226\,m) and extended ($L_{80\%}{\sim}$709\,m) configurations of the 12m ALMA Array. This offers an angular resolution of 0\arcsec.9, and a maximum recoverable scale of 14\arcsec.7. The receiver setup includes four spectral windows and continuous frequency coverage around 89 and 101\,GHz (see Table \ref{table:obspar-ALMA}), with a total bandwidth of 7.6\,GHz.

Project 2018.1.01113.S (PI: Sahai) provides higher frequency observations of V Hya with both the 12\,m Array and the 7\,m Atacama Compact Array (ACA). These observations were used in the analysis by \citet{Sahai2022_VHyaDUDE}, who constrained the physical structure and conditions of the expanding circumstellar disk. While the frequency setups are designed primarily to map \ce{^{12}CO} and \ce{^{13}CO}, the additional spectral windows and bandwidth coverage allow for serendipitous detections of numerous species which we aim to characterize in this work. As shown in Table \ref{table:obspar-ALMA}, observations from this project were performed at Bands 6 and 7, with spatial resolutions 0\arcsec.6 and 0\arcsec.5, respectively. Accompanying ACA observations are available at Band 7 to increase the maximum recoverable scale of emission; however, none were taken for Band 6. \citet{Sahai2022_VHyaDUDE} discussed the flux recovery in this data set, noting that the 12m \ce{^{13}CO} lines experience up to 25\% flux loss, so this effect must be considered when examining Band 6 maps of species with similarly extended distributions.

All observations are located toward $\alpha_\mathrm{J2000}$ = 10$^{\mathrm{h}}$51$^{\mathrm{m}}$37.241$^{\mathrm{s}}$, $\delta_\mathrm{J2000}$ = -21\arcdeg15\arcmin00.28\arcsec; a small deviation in the pointing coordinates of the two ALMA projects was corrected in the imaging process by changing the phase center. Data were newly reduced by these authors using the standard ALMA calibration pipeline, and analysis was performed with the Common Astronomy Software Application \citep[CASA; ][]{CASA2022}. For observations that were performed with multiple array configurations, and overlapping frequency ranges, data were combined in the uv-plane using antenna-specific weighting. Continuum subtraction was performed on visibilities using a first-order fit to line-free channels in each spectral window. Image cubes were then produced with the TCLEAN task using Briggs weighting and a robust parameter of 0.5. For the full bandwidth products (used to create the spectra in Section \ref{section:spectra}), auto-masking was used to identify clean components in the deconvolution process \citep{Kepley2020_automultithresh}. After bright molecular lines were identified and cataloged, their visibilities were re-imaged with an interactive masking procedure to ensure that no artifacts were included in the mask.

Self-calibration was not performed for any of the data sets, as the signal-to-noise of pipeline calibrated data was sufficient for our analysis. The uncertainty in the flux calibration is 5--10\%, as is standard for ALMA observations \citep{Cortes2020_ALMAtechnical}.

\section{Results}
\label{section:results}
\subsection{Spectra and detected transitions}
\label{section:spectra}

Spectra were extracted from image cubes using a 4"$\times$6" aperture (aligned with the major axis of the DUDE; position angle ${\sim}$ -$5^{\circ}$). While this is smaller than the observed emission of the most extended molecules (Section \ref{section:maps}), we adopt this aperture to increase the sensitivity to weaker lines, which we find are often present in more compact distributions toward V Hya. For the majority of detected molecules, this aperture is large enough to include all of the observed flux. All data were shifted to rest frequency units using the system velocity of V Hya, $v_{\mathrm{lsr}}=-17.4$\,\kms.

The continuum-subtracted spectra of V Hya are shown in Figure \ref{fig:specs}, including labels for all species contributing emission lines. This includes all spectral ranges shown in Table \ref{table:obspar-ALMA}, except for the narrower 345.6--346\,GHz window which is dominated only by major transitions of \ce{^{12}CO} and \ce{HC3N}. We find that the spectrum of V Hya is relatively line-poor at Band 3, with only seven lines appearing across almost 8\,GHz of bandwidth. This is in contrast to other C-rich AGB envelopes, which typically show rich spectra at these wavelengths and sensitivities (e.g. see Fig.\ 4 in \citet{Unnikrishnan2024_charting}). The most notable molecules that are not present in these regions are \ce{C3N} and \ce{SiS}, whose transitions in the case of IRC+10216 are equally as bright as \ce{HC5N} and \ce{HNC} at Band 3 \citep{Tuo2024_IRC_survey}. At Bands 6 and 7 however, the spectrum of V Hya is much richer, as together they account for over 90\% of the observed lines.

\begin{figure*}[h!]
    \centering
    \includegraphics[width=\linewidth]{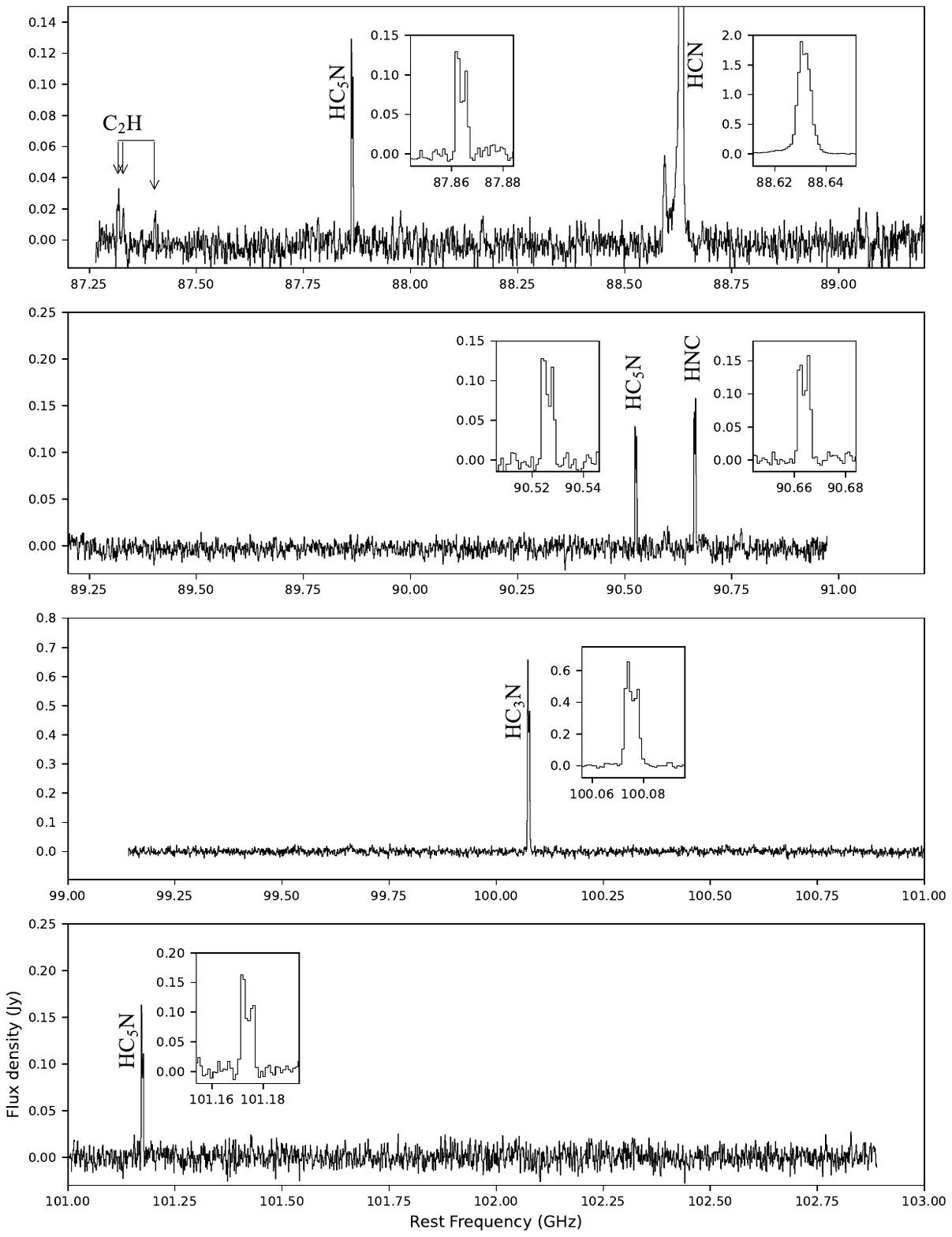}
    \caption{}
\end{figure*}
\begin{figure*}[h!]
    \ContinuedFloat
    \centering
    \includegraphics[width=\linewidth]{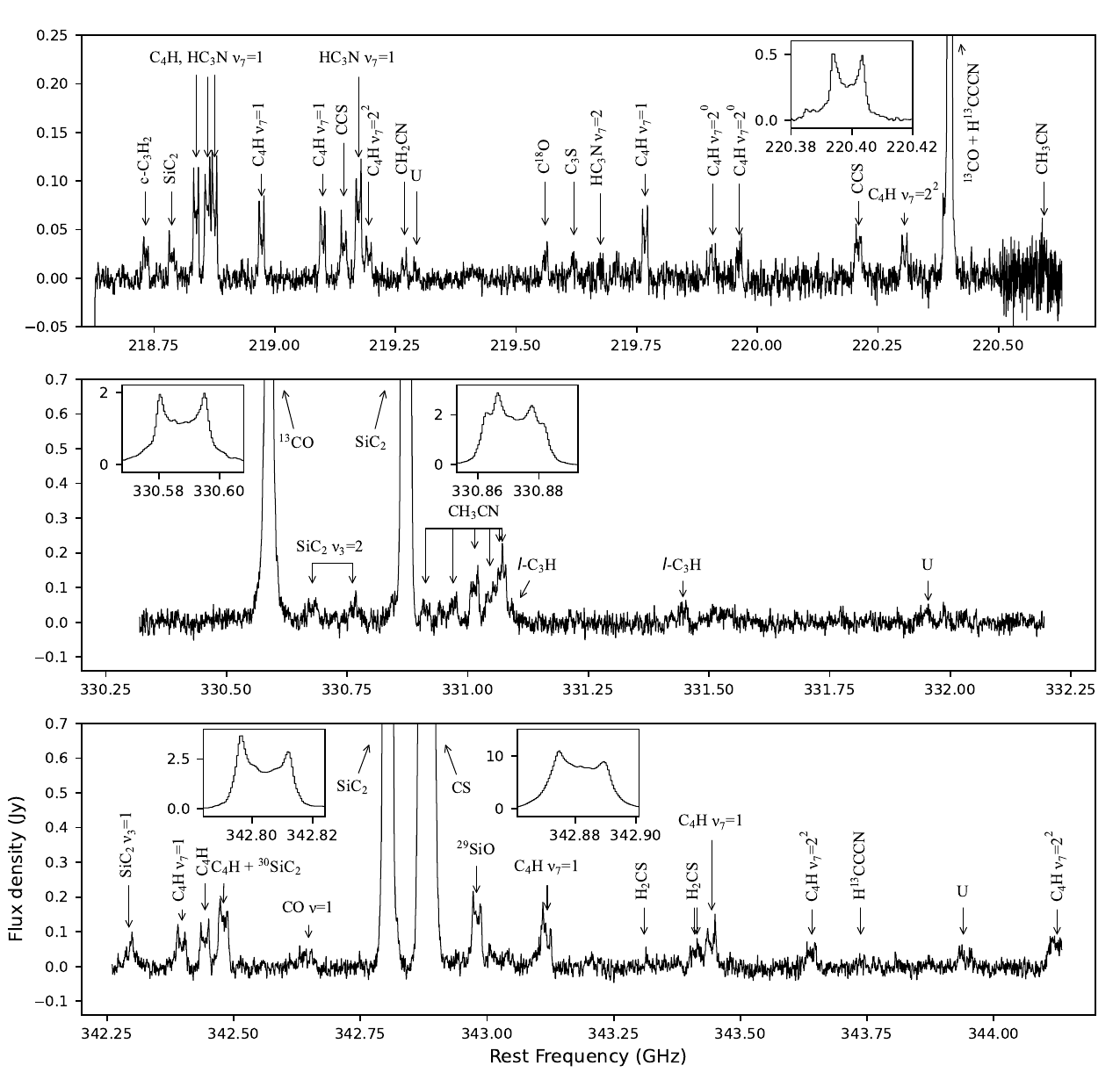}
    \caption{Full spectrum of V Hya obtained from the ALMA observations. Spectra were extracted using a 4"$\times$6" elliptical aperture on the position of V Hya, and shifted to rest frequency using the system velocity $v_{\mathrm{lsr}}=-17.4$\,\kms. All identified transitions are labeled, and insets are provided for bright lines. Not included here is a separate spectral window covering 345.6--346\,GHz, which contains only the \ce{^{12}CO} $J=3-2$ and \ce{HC3N} $J=38-37$ lines.}
    \label{fig:specs}
\end{figure*}
\clearpage

\startlongtable
\begin{deluxetable*}{lccccccc}
    \tablecaption{List of all detected transitions}
    \tablewidth{\columnwidth}
    \tablehead{
    \colhead{Rest Frequency} & \colhead{Molecule} & \colhead{Transition}  &\colhead{E$_{\mathrm{up}}$} &\colhead{$S_{ij}\mu^2$}  & \colhead{$v{_\mathrm{exp}}$} & \colhead{$\int Sdv$} & \colhead{$\sigma^a$}\\ (MHz) &  & & (K)& (D$^2$) & \kms & (Jy \kms) & (Jy \kms)} 
    \startdata
\phantom{0}87 316.83(310) & CCH & $N=1-0$, $J=3/2-1/2$, $F=2-1$ & 4.193 & 1.425 & 8.4 $\pm$ 3.0 & 0.4548 & 0.106 \\
\phantom{0}87 327.55(75) & CCH & $N=1-0$, $J=3/2-1/2$, $F=1-0$ & 4.193 & 0.7095 & 8.34 $\pm$ 4.0 & 0.2532 & 0.106 \\
\phantom{0}87 402.00 & CCH & $N=1-0$, $J=1/2-1/2$, $F=1-1$ & 4.193 & 0.7095 &  & -- &  \\
\phantom{0}87 863.59(5) & HC$_{5}$N & $J=33-32$ & 71.69 & 1856.0 & 7.91 $\pm$ 0.2 & 1.75 & 0.116 \\
\phantom{0}88 593.39(33)$^b$ & HCN & $J=1-0$ & 4.254 & 26.82 &12.4 $\pm$ 1.0 & 1.351 & 0.116 \\
\phantom{0}88 630.96(8) & HCN & $J=1-0$\phantom{$^b$} & 4.254 & 26.82 &9.74 $\pm$ 0.3 & 38.61$^c$ & 0.116 \\
\phantom{0}90 525.86(61) & HC$_{5}$N & $J=34-33$ & 76.03 & 1912.0 & 8.04 $\pm$ 1.0 & 1.881 & 0.108 \\
\phantom{0}90 663.42(2) & HNC & $J=1-0$ & 4.351 & 9.302 & 7.68 $\pm$ 0.1 & 2.377 & 0.108 \\
100 076.25(3) & HC$_{3}$N & $J=11-10$ & 28.82 & 153.2 & 7.84 $\pm$ 0.2 & 9.511 & 0.141 \\
101 174.58(5) & HC$_{5}$N & $J=38-27$ & 94.69 & 2138.0 & 7.63 $\pm$ 0.2 & 2.118 & 0.137 \\
 & & & & & & &  \\
218 732.70(16) & c-C$_{3}$H$_{2}$ & $J_{K_a,K_c}=7_{1,6}-7_{0,7}$ , $7_{2,6}-7_{1,7} $ & 61.17 & 44.0 & 7.88 $\pm$ 0.3 & 0.4571 & 0.11 \\
218 786.12(16) & SiC$_{2}$ & $J_{K_a,K_c}=11_{2,10}-11_{0,11}$ & 81.85 & 1.004 & 8.36 $\pm$ 0.3 & 0.4518 & 0.11 \\
218 836.59(6) & C$_{4}$H & $N=23-22$, $J=47/2-45/2$ & 126.0 & 207.1 & 8.01 $\pm$ 0.1 & 1.293 & 0.11 \\
218 860.80 & HC$_{3}$N $\nu_7$=1 & $J=24-23$, $l=1e$ & 452.1 & 332.0 &  & -- &  \\
218 875.35 & C$_{4}$H & $N=23-22$, $J=45/2-43/2$ & 126.0 & 198.2 &  & -- &  \\
218 972.30(6) & C$_{4}$H $\nu_7$=1 & $J=45/2-43/2$, $l=f$ & 372.8 & 232.0 & 8.01 $\pm$ 0.1 & 0.8833 & 0.11 \\
219 098.98(7) & C$_{4}$H $\nu_7$=1 & $J=47/2-45/2$, $l=e$ & 371.4 & 242.0 & 7.54 $\pm$ 0.2 & 0.8625 & 0.11 \\
219 142.49(14) & CCS & $N=17-16$, $J=16-15$ & 100.1 & 132.3 & 8.17 $\pm$ 0.4 & 0.7707 & 0.11 \\
219 173.35(10) & HC$_{3}$N $\nu_7$=1 & $J=24-23$, $l=1f$ & 452.3 & 332.0 & 8.35 $\pm$ 0.3 & 1.594 & 0.11 \\
219 194.46(15) & C$_{4}$H $\nu_7$=$2^2$ & $J=47/2-45/2$ & 617.1 & 394.1 & 7.76 $\pm$ 0.6 & 0.5624 & 0.11 \\
219 267.89(11) & CH$_{2}$CN & $J=21/2-19/2$ & 76.35 & 381.8 & 7.09 $\pm$ 0.2 & 0.236 & 0.11 \\
219 293.90(19) & U &  -- &  -- &  -- & 8.44 $\pm$ 0.3 & 0.1342 & 0.11 \\
219 560.00(17) & C$^{18}$O & $J=2-1$ & 15.81 & 0.0244 & 7.39 $\pm$ 0.4 & 0.3421 & 0.11 \\
219 620.10(30) & C$_{3}$S & $J=38-37$ & 205.6 & 521.3 & 7.9\phantom{ $\pm$ 0.0} & 0.2364 & 0.11 \\
219 675.11 & HC$_{3}$N $\nu_7$=2 & $J=24-23$, $l=0$ & 773.5 & 331.1 &  & -- &  \\
219 766.95(7) & C$_{4}$H $\nu_7$=1 & $J=47/2-45/2$, $l=f$ & 371.8 & 232.0 & 8.02 $\pm$ 0.1 & 0.8241 & 0.11 \\
219 908.65(26) & C$_{4}$H $\nu_7$=$2^0$ & $J=47/2-45/2$, $N=23-22$ & 662.9 & 242.7 & 5.28 $\pm$ 2.0 & 0.4298 & 0.11 \\
219 961.44(18) & C$_{4}$H $\nu_7$=$2^0$ & $J=45/2-43/2$, $N=23-22$ & 662.9 & 232.3 & 8.54 $\pm$ 0.3 & 0.3844 & 0.11 \\
220 208.88(24) & CCS & $N=17-16$, $J=17-16$ & 103.8 & 140.5 & 6.33 $\pm$ 2.0 & 0.8523 & 0.11 \\
220 304.86(14) & C$_{4}$H $\nu_7$=$2^2$ & $J=45/2-43/2$ & 615.6 & 410.4 & 7.42 $\pm$ 0.6 & 0.5849 & 0.11 \\
220 390.20 & H$^{13}$CCCN & $J=25-24$ & 137.5 & 348.1 &  & -- &  \\
220 398.68 & $^{13}$CO & $J=2-1$ & 15.87 & 0.04869 &  & -- &  \\
220 594.32(19) & CH$_{3}$CN & $J_K=12_6-11_6$ & 325.9 & 381.5 & 8.83 $\pm$ 0.3 & 0.3082 & 0.264 \\
 & & & & & & &  \\
330 587.49(15) & $^{13}$CO & $J=3-2$ & 31.73 & 0.07297 & 7.52 $\pm$ 0.4 & 27.39 & 0.282 \\
330 677.46(53) & SiC$_{2}$ $\nu_3$=2 & $J_{K_a,K_c}=13_{2,14}-12_{2,13}$ & 645.1 & 83.94 & 6.71 $\pm$ 2.0 & 1.032 & 0.282 \\
330 760.82(35) & SiC$_{2}$ $\nu_3$=2 & $J_{K_a,K_c}=16_{0,16}-15_{0,15}$ & 646.3 & 90.96 & 7.9\phantom{ $\pm$ 0.0} & 0.7261 & 0.282 \\
330 872.01(12) & SiC$_{2}$ & $J_{K_a,K_c}=14_{6,9}-13_{6,8}$ & 189.1 & 65.47 & 7.48 $\pm$ 0.4 & 44.45 & 0.282 \\
330 913.37(37) & CH$_{3}$CN & $J_K=18_5-17_5$ & 329.5 & 352.1 & 8.13 $\pm$ 0.5 & 0.6962 & 0.282 \\
330 969.07(41) & CH$_{3}$CN & $J_K=18_4-17_4$ & 265.2 & 362.7 & 7.98 $\pm$ 0.7 & 0.905 & 0.282 \\
331 013.44(18) & CH$_{3}$CN & $J_K=18_3-17_3$ & 215.2 & 741.9 & 8.43 $\pm$ 0.3 & 1.985 & 0.282 \\
331 046.10 & CH$_{3}$CN & $J_K=18_2-17_2$ & 179.5 & 376.9 &  & -- &  \\
331 065.18 & CH$_{3}$CN & $J_K=18_1-17_1$ & 158.1 & 380.4 &  & -- &  \\
331 071.54 & CH$_{3}$CN & $J_K=18_0-17_0$ & 151.0 & 381.6 &  & -- &  \\
331 084.76 & $l$-C$_3$H & $J=29/2-27/2$, $l=f$ & 142.7 & 360.9 &  & -- &  \\
331 445.88(17) & $l$-C$_3$H & $J=29/2-27/2$, $l=e$ & 142.9 & 360.9 & 9.69 $\pm$ 0.2 & 0.622 & 0.282 \\
331 954.00 & U &  -- &  -- &  -- &  & -- & 0.282 \\
 & & & & & & &  \\
342 292.93(43) & SiC$_{2}$ $\nu_3$=1 & $J_{K_a,K_c}=15_{9,6}-14_{9,5}$ & 578.2 & 110.0 & 7.9\phantom{ $\pm$ 0.0} & 0.8341 & 0.235 \\
342 396.94(20) & C$_{4}$H $\nu_7$=1 & $J=73/2-71/2$, $l=e$ & 549.5 & 376.7 & 7.87 $\pm$ 0.4 & 1.434 & 0.235 \\
342 442.95(11) & C$_{4}$H & $N=36-35$, $J=73/2-71/2$ & 304.1 & 317.4 & 8.1 $\pm$ 0.2 & 1.583 & 0.235 \\
342 479.35 & $^{30}$SiC$_{2}$ & $J_{K_a,K_c}=15_{8,8}-14_{8,7}$ & 257.4 & 61.48 &  & -- &  \\
342 481.36 & C$_{4}$H & $N=36-35$, $J=71/2-69/2$ & 304.2 & 308.6 &  & -- &  \\
342 646.31(32) & CO $\nu$=1 & $J=3-2$ & 3117.0 & 0.02164 & 7.9\phantom{ $\pm$ 0.0} & 0.5096 & 0.235 \\
342 804.47(7) & SiC$_{2}$ & $J_{K_a,K_c}=15_{2,14}-14_{2,13}$ & 141.4 & 83.93 & 8.26 $\pm$ 0.1 & 44.64 & 0.235 \\
342 882.22(10) & CS & $J=7-6$ & 65.83 & 26.75 & 7.79 $\pm$ 0.3 & 178.4 & 0.235 \\
342 980.09(16) & $^{29}$SiO & $J=8-7$ & 74.08 & 76.78 & 7.87 $\pm$ 0.4 & 3.092 & 0.235 \\
343 118.11(42) & C$_{4}$H $\nu_7$=1 & $J=71/2-69/2$, $l=f$ & 551.2 & 366.5 & 7.88 $\pm$ 0.9 & 2.139 & 0.235 \\
343 309.83 & H$_{2}$CS & $J_{K_a,K_c}=10_{4,7}-9_{4,6}$ & 301.1 & 22.84 &  & -- &  \\
343 322.09 & H$_{2}$CS & $J_{K_a,K_c}=10_{2,9}-9_{2,8}$ & 143.3 & 26.11 &  & -- &  \\
343 409.17(38) & H$_{2}$CS & $J_{K_a,K_c}=10_{3,8}-9_{3,7}$ & 209.1 & 74.24 & 7.9\phantom{ $\pm$ 0.0} & 0.7369 & 0.235 \\
343 414.15 & H$_{2}$CS & $J_{K_a,K_c}=10_{3,7}-9_{3,6}$ & 209.1 & 74.24 &  & -- &  \\
343 442.43(18) & C$_{4}$H $\nu_7$=1 & $J=73/2-71/2$, $l=f$ & 550.4 & 376.7 & 7.95 $\pm$ 0.4 & 1.673 & 0.235 \\
343 640.05(19) & C$_{4}$H $\nu_7$=$2^2$ & $J=73/2-71/2$ & 795.7 & 624.3 & 8.8 $\pm$ 0.2 & 0.8023 & 0.235 \\
344 123.63(35) & C$_{4}$H $\nu_7$=$2^2$ & $J=71/2-69/2$ & 794.7 & 641.4 & 7.9\phantom{ $\pm$ 0.0} & 1.133 & 0.235 \\
343 737.40 & H$^{13}$CCCN & $J=39-38$ & 330.0 & 543.1 &  & -- &  \\
343 942.00(74) & U &  -- &  -- &  -- & 10.9 $\pm$ 1.0 & 0.6825 & 0.235 \\
 & & & & & & &  \\
345 608.26(5) & HC$_{3}$N & $J=38-37$ & 323.5 & 529.1 & 8.4 $\pm$ 0.1 & 25.06 & 0.576 \\
345 795.99 & CO & $J=3-2$ & 33.19 & 0.03631 &  & -- &  \\
\enddata   
    \tablecomments{Spectroscopic data were obtained from Cologne Database of Molecular Spectroscopy (CDMS) \citep{2005JMoSt.742..215M}, except in the case of \ce{CH3CN}, for which the Jet Propulsion Laboratory catalog was used \citep{Pickett_JPL}. For blended lines and those with peak flux under $3\sigma$, no fit was performed, and the catalog frequency is shown with no uncertainty. In the case of weak lines that initially failed to converge, the line width was fixed to 7.9\,\kms and no error for that parameter is listed. $^a$Reported rms is given over 17\,\kms line width. $^b$Redshifted component of HCN emission, separate fit performed. $^c$Gaussian profile used for line fit, listed expansion velocity is half the FWHM.}
    \label{table:lines}
\end{deluxetable*}

\begin{deluxetable}{cccc}
    \tablecaption{Detected molecules, isotopologues, and vibrationally excited states toward V Hya}
    \tablewidth{\linewidth}
    \tablehead{
    \colhead{Species} & \colhead{No.\ lines}  &\colhead{E$_{\mathrm{up}}$ Range (K)} &\colhead{Vib.\ States}} 
    \startdata
    \ce{^{12}CO} & 2 & 33--3100 & $\nu=1$ \\
    \ce{^{13}CO} & 2 & 16--32 &  \\
    \ce{C^{18}O} & 1 & 16 &  \\
    \ce{HCN} & 1 & 4.2 &  \\
    \ce{SiC2} & 6 & 82--650 & $\nu_3=2$, $\nu_3=1$ \\
    \ce{^{30}SiC2} & 1 & 260 &  \\
    \ce{^{29}SiO} & 1 & 74 &  \\
    \ce{CS} & 1 & 66 &  \\
    \ce{HNC} & 1 & 4.3 &  \\
    \ce{CCH} & 3 & 4.2 &  \\
    \ce{l-C3H} & 2 & 140 &  \\
    \ce{C4H} & 15 & 130--800 & $\nu_7=1$, $\nu_7=2^2$, $\nu_7=2^0$ \\
    \ce{c-C3H2} & 1 & 61 &  \\
    \ce{HC3N} & 4 & 29--770 & $\nu_7=1$, $\nu_7=2$ \\
    \ce{H^{13}CCCN} & 1 & 138 &  \\
    \ce{HC5N} & 3 & 72--95 &  \\
    \ce{CCS} & 2 & 100 &  \\
    \ce{C3S} & 1 & 200 &  \\
    \ce{CH3CN} & 7 & 150--380 &  \\
    \ce{CH2CN} & 2 & 76 &  \\
    \ce{H2CS} & 2 & 143--300 &  \\
    \enddata
    \tablecomments{If no vibrational state is listed, only the ground state is detected. No species are found in only vibrationally excited lines.}
\label{table:mol_count}
\end{deluxetable}

Table 2 lists the identified transitions in Fig.\ \ref{fig:specs} with their quantum numbers and spectroscopic constants (upper state energy, line strength). In total, we detect emission from 16 unique molecules, 20 including isotopologues. The number of lines detected for each species, as well as detected energy states and vibrational modes, are summarized in Table 3. For four molecules (\ce{^{12}CO}, \ce{HC3N}, \ce{C4H}, and \ce{SiC2}), we see rotational lines from excited vibrational modes, all other transitions are in the ground vibrational state. A full discussion of each molecule, along with its observed spatial distribution and history of study in C-rich AGB outflows is provided in Section \ref{section:mol_disc}.

\begin{figure}[h!]
    \centering
    \includegraphics[width=\linewidth]{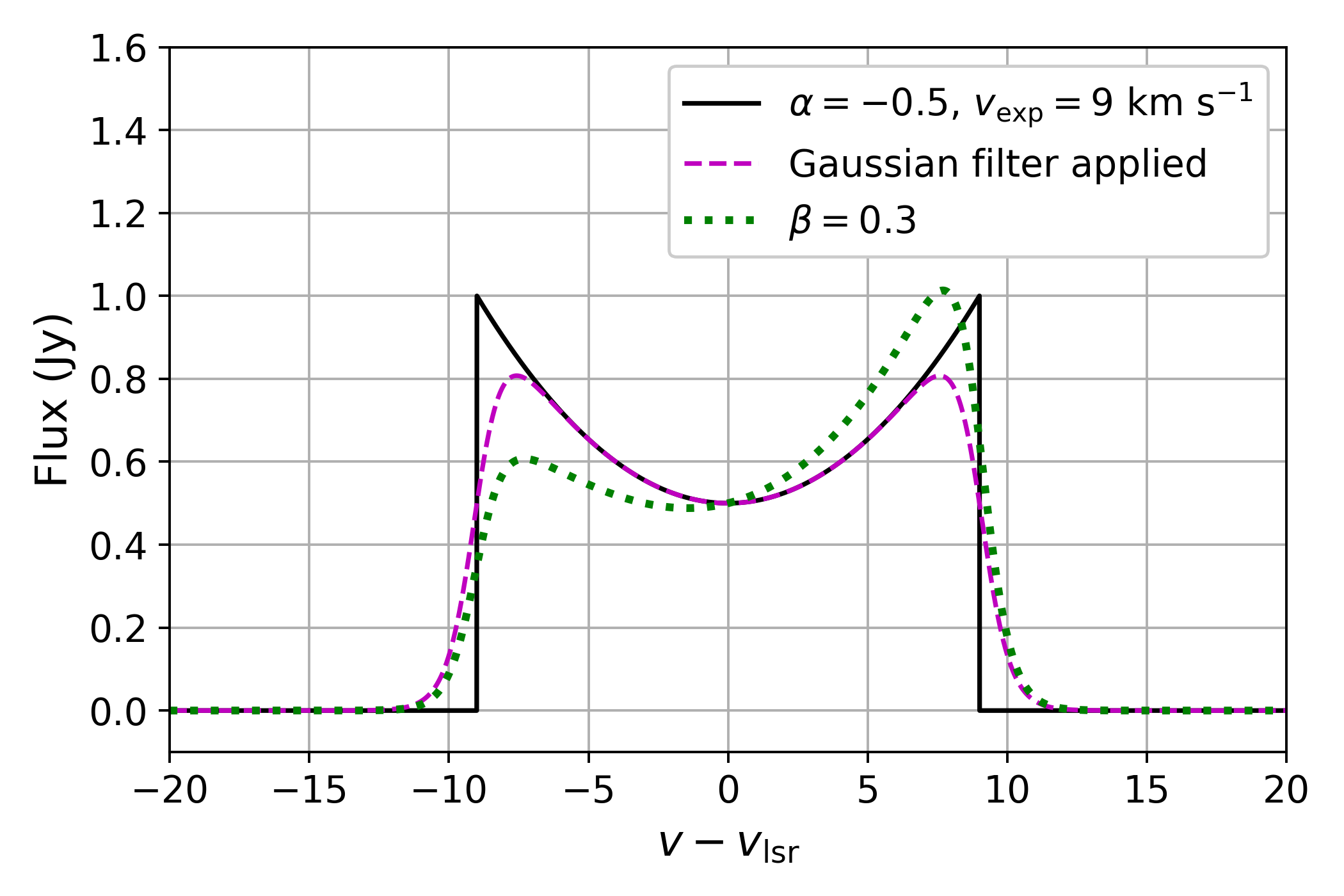}
    \caption{Demonstration of the parametric model used to fit individual lines. Solid line depicts the original double-peaked model of \citet{Wannier1990_AGBlines} with $S_{\mathrm{pk}}=1$\,Jy, the dashed line represents the result of applying a gaussian top hat filter to model line wings, and the dotted line shows the effect of using a positive non-zero value for the asymmetry parameter $\beta$ (Eq. \ref{eq:line_prof}). }
    \label{fig:fit_params}
\end{figure}

The line profiles seen in Fig. \ref{fig:specs} are nearly all double-peaked, as was observed in \citet{Sahai2022_VHyaDUDE}. The only exception to this is the $J=1-0$ transition of \ce{HCN}, which shows a central peak and parabolic shape, indicating that this emission is optically thick \citep{Smith2015_molsurvey_AGB_PNe}. To measure the position and integrated flux of detected lines, we fit them individually with an empirical intensity profile that reproduces the shape of emission. Following the formalism of \citet{Wannier1990_AGBlines}, the profile consists of a central parabolic shape that peaks at $\pm v_{\mathrm{exp}}$, the radial expansion velocity of the DUDE:

\begin{equation}
\label{eq:line_prof}
    S(v)=S_{\mathrm{pk}}\times\left(1-\alpha\left[1-\left(\frac{v-v_{\mathrm{lsr}}}{v_{\mathrm{exp}}}\right)^2\right]\right)
\end{equation}
where $S_{\mathrm{pk}}$ is the peak flux, $v_{\mathrm{lsr}}$ is the source velocity, and $\alpha$ is a shaping parameter ranging from zero (flat-topped profile) to one (intensity minimum of 0 at $v_{\mathrm{exp}}$). Since this function is only applicable for $|v-v_{\mathrm{lsr}}|<v_{\mathrm{exp}}$, a gaussian top hat filter is applied to reproduce the line wings (Fig.\ \ref{fig:fit_params}). We also must include an asymmetry parameter which allows for the two peaks to have different intensities, as is observed in many transitions (e.g.\ \ce{CS}, \ce{SiC2}, \ce{HC5N}). This is incorporated by scaling the profile in Eq.\ \ref{eq:line_prof} with a linear function $\beta$ that can range from $-1$ (all flux in blue-shifted peak) to zero (symmetric line) to $+1$ (all red-shifted):

\begin{equation}
    S_{\mathrm{asym}}(v)=S(v)\times\left(1+\beta\left[\frac{v-v_{\mathrm{lsr}}}{v_{\mathrm{exp}}}\right]\right)
\end{equation}

\begin{figure}[h!]
    \centering
    \includegraphics[width=\linewidth]{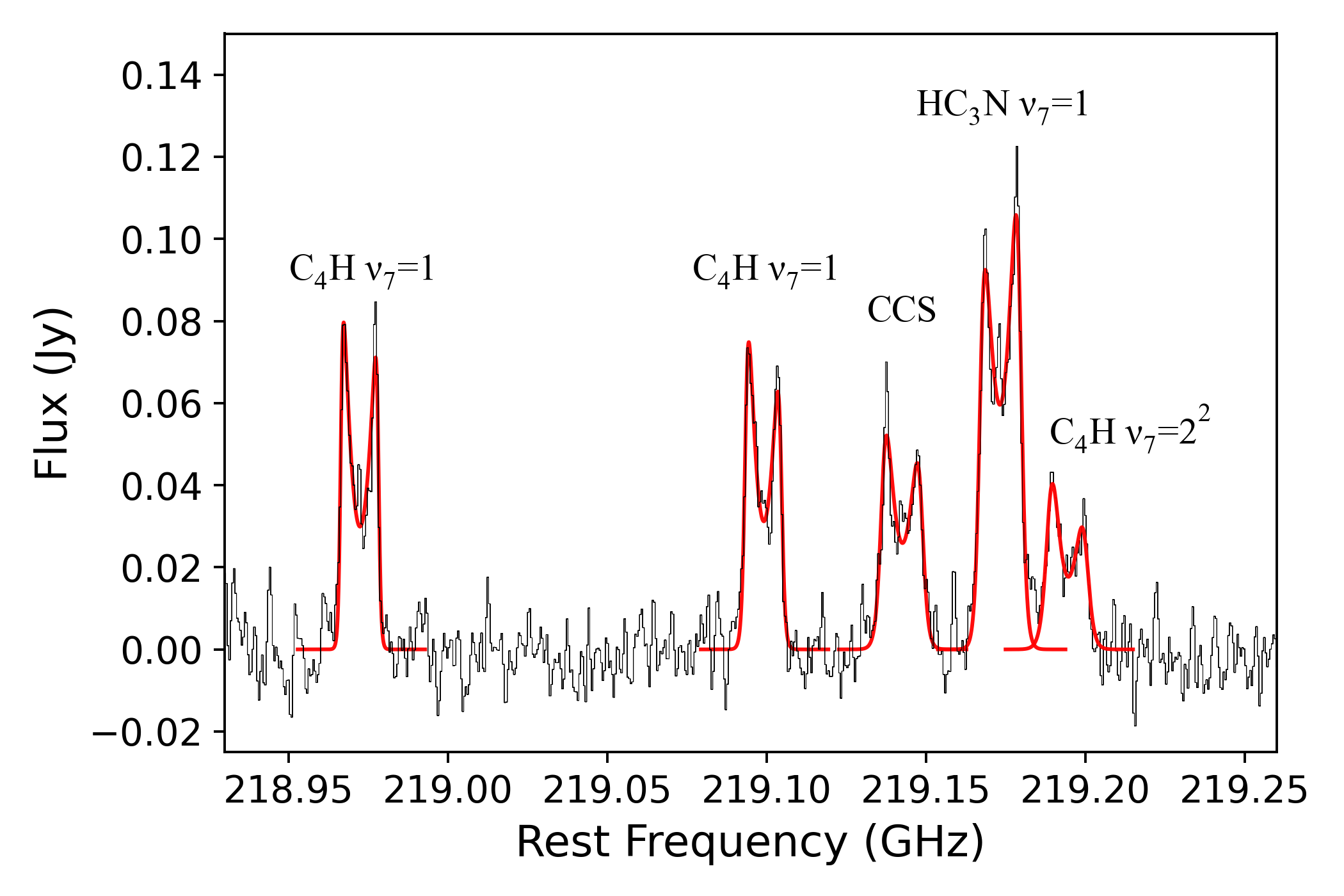}
    \caption{Individual line-fitting results for a group of transitions at Band 6. The observed ALMA spectrum of V Hya is shown in black, while the model fits are overlaid in red. The best-fit central frequencies, expansion velocities, and resulting integrated line fluxes are listed in Table \ref{table:lines}.}
    \label{fig:fitting}
\end{figure}

In total, we have five variables ($S_{\mathrm{pk}}$, $v_{\mathrm{lsr}}$, $v_{\mathrm{exp}}$, $\alpha$, $\beta$ ) determining the line shape which we fit with a least-squares minimization. We apply this to every unblended transition that is detected above $3\sigma$, and integrate the emission models to obtain the integrated line fluxes. This value, along with the best-fit central frequency, and expansion velocity are listed in Table 2, and a sample of these fits is shown in Fig.\ \ref{fig:fitting}.

We find that almost all transitions are well-fit with an expansion velocity of ${\sim}$8\,\kms, which is consistent with the physical model of the DUDE proposed in \citet{Sahai2022_VHyaDUDE}. We also find no evidence of lines that solely arise from the innermost regions of the outflow ($<$200\,au), where gas is still being accelerated and thus yields narrower profiles \citep{Decin2018_highlowMLR_alma}. This is supported in our spatial analysis of these data (Section \ref{section:maps}), as all molecular emission is found to be at least partially resolved by the ALMA beam.

We report three unidentified lines centered at rest frequencies of 219\,294\,MHz, 331954\,MHz, and 343\,940\,MHz. Another feature which could be classified as a U line appears very close to the $J=1-0$ transition of \ce{HCN}. We instead interpret this as highly redshifted component of \ce{HCN}, where it would correspond to a velocity of +127\,\kms. Though this is very fast for circumstellar envelopes, it is remarkably close to the redshifted high-velocity bullet observed in \ce{^{12}CO} by \citet{Sahai2022_VHyaDUDE}. It also exhibits a small westward offset in position that is associated with this clump. Because of this, we conclude that the emission is contributed by \ce{HCN} in a dense, high-velocity clump ejected by V Hya, and fit it with a separate profile listed in Table 2.

\subsection{Emission maps}
\label{section:maps}

\begin{figure*}[t!]
    \centering
    \includegraphics[width=\linewidth,trim=0.1cm 0.1cm 0.5cm 1.4cm,clip]{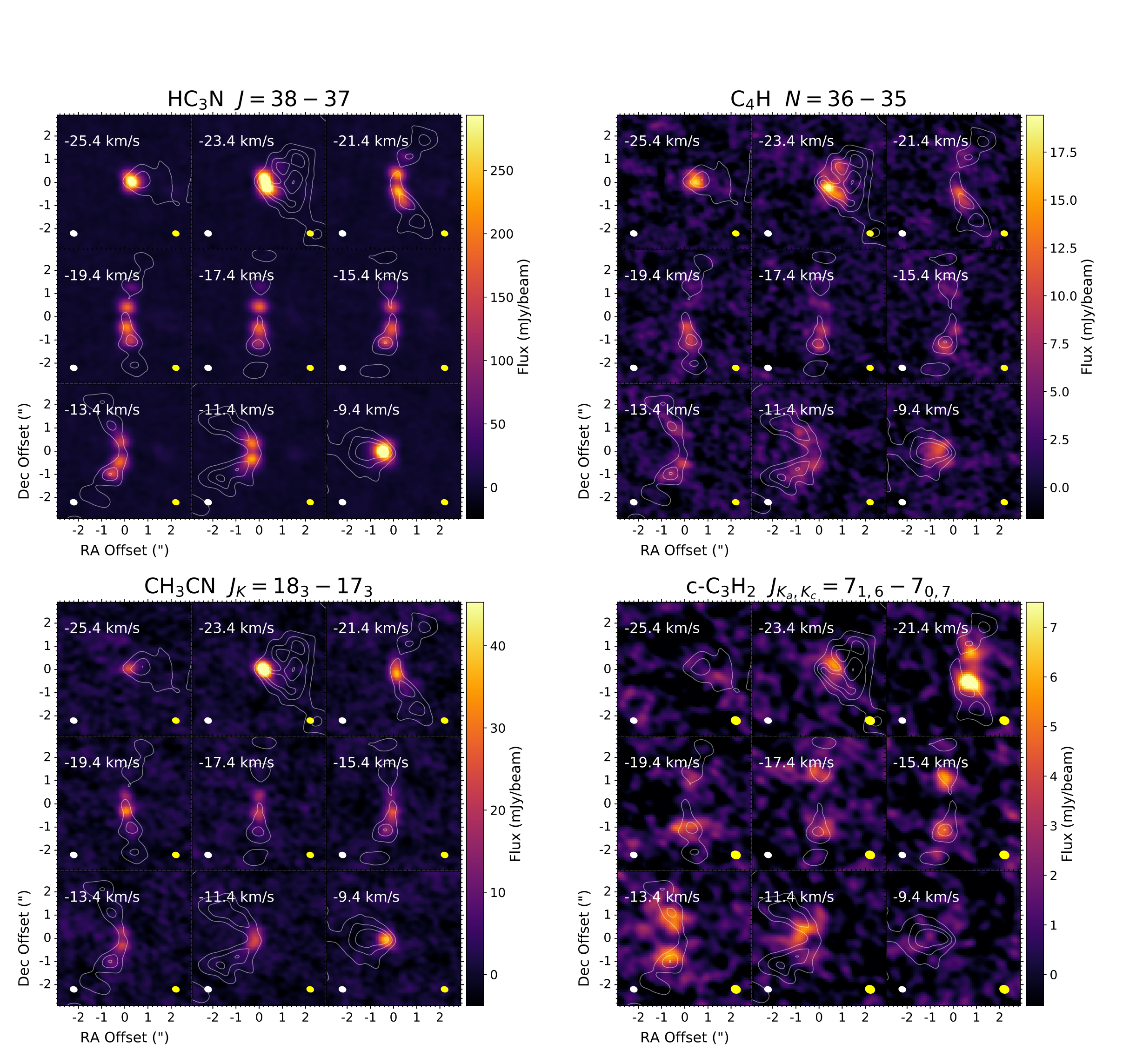}
    \caption{Channel maps of selected C-bearing molecular transitions. All images are overlaid with corresponding channel maps from \ce{^{13}CO} $J=3-2$ in white at 20\%, 40\%, and 60\% the peak flux of that image (0.166\,Jy). White ellipses denote the synthesized beam for the \ce{^{13}CO} window, while yellow ellipses indicate the shape of the beam for the mapped molecule.}
    \label{fig:chmaps}
\end{figure*}

To explore the spatial morphologies of chemical products in the DUDE, we first create channel maps for all unblended transitions above a $5\sigma$ intensity threshold. The channel maps span a range of 50\,\kms and are centered on the system velocity of V Hya. A sample of these channel maps, chosen to demonstrate the range of regions in the disk sampled by hydrocarbon species, is shown in Fig.\ \ref{fig:chmaps}. The transitions mapped here (\ce{HC3N}, \ce{C4H}, c-\ce{C3H2}, and \ce{CH3CN}) show the expected velocity pattern of an expanding disk inclined (roughly) along the N-S axis, and mostly appear to be spatially coincident with \ce{^{13}CO}, and thus the density structure of the outflow. In our discussion of these maps, we adopt the nomenclature of \citet{Sahai2022_VHyaDUDE}, who labeled the major density rings at 560, and 1040, and 1740\,au as R1, R2, and R3, respectively. In the $-17.4$\,\kms velocity channel seen in Fig.\ \ref{fig:chmaps}, the positions of the R1 and R2 structures can be seen as the two peaks in the N-S direction at distances of 1\arcsec and 2.5\arcsec from the image center. 

\citet{Sahai2022_VHyaDUDE} also identified a central ``R0" ring from the distribution of the \ce{HC3N} $J=38-37$ transition, apparent in Fig.\ \ref{fig:chmaps} as well. They note that this should be considered separately from the other rings in the DUDE, as it is not reproduced in \ce{^{12}CO} or \ce{^{13}CO}, and therefore could result from chemistry in this region of the DUDE. We now see in Fig.\ \ref{fig:chmaps} that this R0 ring is also observed in \ce{C4H}, and potentially \ce{CH3CN} as well. In our calculation of abundances (Section \ref{section:disc}), we will discuss whether the R0 structure is actually a region of increased chemistry producing these daughter species, or rather a location where conditions are energetically favorable for the mapped lines to be excited.

\begin{figure*}[h]
    \centering
    \includegraphics[width=\linewidth,trim=0.5cm 1.2cm 0cm 0.2cm,clip]{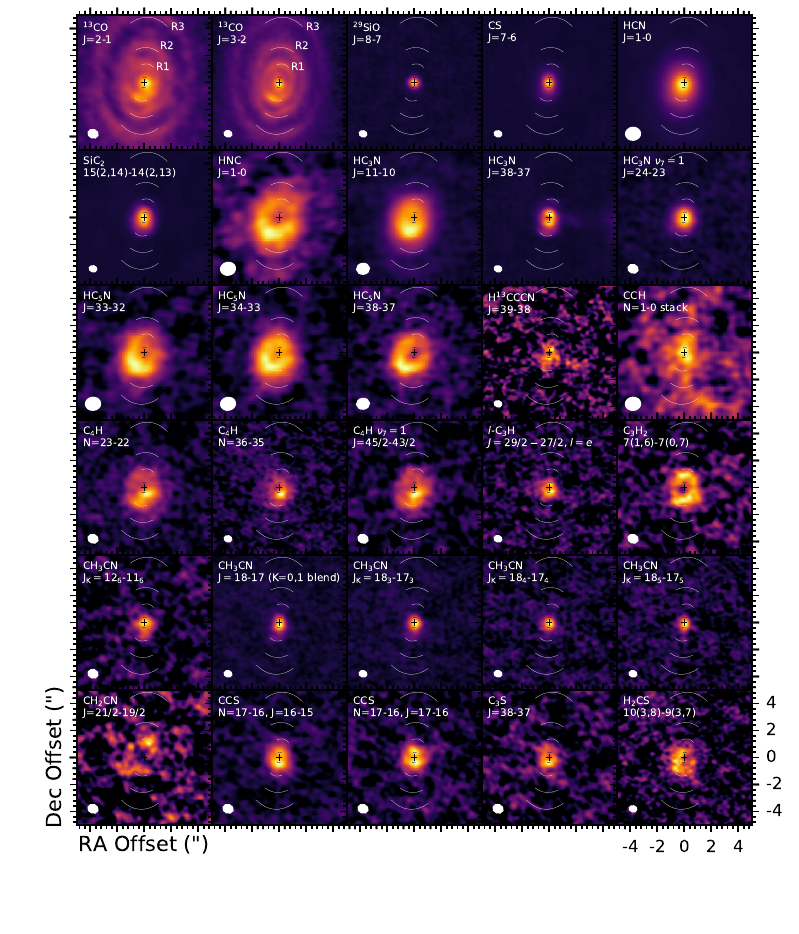}
    \caption{Integrated intensity maps of the brightest transitions detected for all species. Ring structures R1, R2, and R3 are denoted by white arcs in each panel, and the synthesized beams are shown in the bottom-left corners. The compact ring R0 at 160\,au is shown in black arcs surrounding the position of V Hya (black cross). The color scale in each image is adjusted to best show emission sub-structure. Maps were integrated over a width of 21\,\kms centered on the system velocity of V Hya.}
    \label{fig:maps}
\end{figure*}

After image cubes were created for each transition, we then made velocity-integrated intensity maps (moment zero) to investigate the total distribution of emission for each transition. A selection of these maps is shown in Figure \ref{fig:maps}, including the positions of R0, R1, R2, and R3. Most species show ring-like patterns tracing the general structure of the DUDE, and often exhibit intensity minima at the position of V Hya. The exceptions to this are lines with centrally peaked emission maps including parent molecules \ce{SiO}, \ce{CS}, \ce{HCN}, \ce{SiC2}, and peculiarly, the more complex species \ce{CH3CN}.

The vast majority of flux for even the most extended species comes from within the R2 ring. And for lines detected at Bands 6 and 7, emission is primarily constrained within R1, and often traces the innermost ring R0 at 160\,au. In general, the lines observed at Band 3 show a larger area of emission, and a farther radial intensity peak than those at higher frequencies, with \ce{HNC} $J=1-0$ and \ce{HC3N} $J=11-10$ being the only species with emission detected at R3. This is typical for observations of AGB stars, where low-lying rotational states tend to be populated more in cooler regions of the envelope \citep{Agundez2017_IRC_carbonchains,Siebert2022_IRC_HC3N}.

Due to the inclination of the source we expect to see a systematic azimuthal brightness asymmetry, i.e., even if the DUDE was a uniform disk, it  would appear brighter along the N--S direction than along the E--W direction. Interestingly though, in Fig.\ \ref{fig:maps} we also find departures from this expected pattern in the spatial distributions of the most compact molecules. The clearest example of this is \ce{C4H}, which shows brighter emission on the southern side of R0 in every transition we observed (this is also apparent in the channel maps in the top right panel of Fig.\ \ref{fig:chmaps}). The same behavior is also observed in \ce{C3S} and to a lesser extent in \ce{CCS}, and it is notably absent from the $J=38-37$ line of \ce{HC3N}. This effect is different from the anisotropy noted in the R1 ring by \citet{Sahai2022_VHyaDUDE} (which can also be seen in panels of Fig.\ \ref{fig:maps} with emission at R1), as it is not observed in the density tracing \ce{^{13}CO} lines, and it is only specific to these molecules. This points to these brightness enhancements being driven by asymmetric chemistry in the inner regions of the DUDE.

\subsection{Kinematic structure and radial profiles}
\label{section:profs}

\begin{figure}[h]
    \centering
    \includegraphics[width=\linewidth]{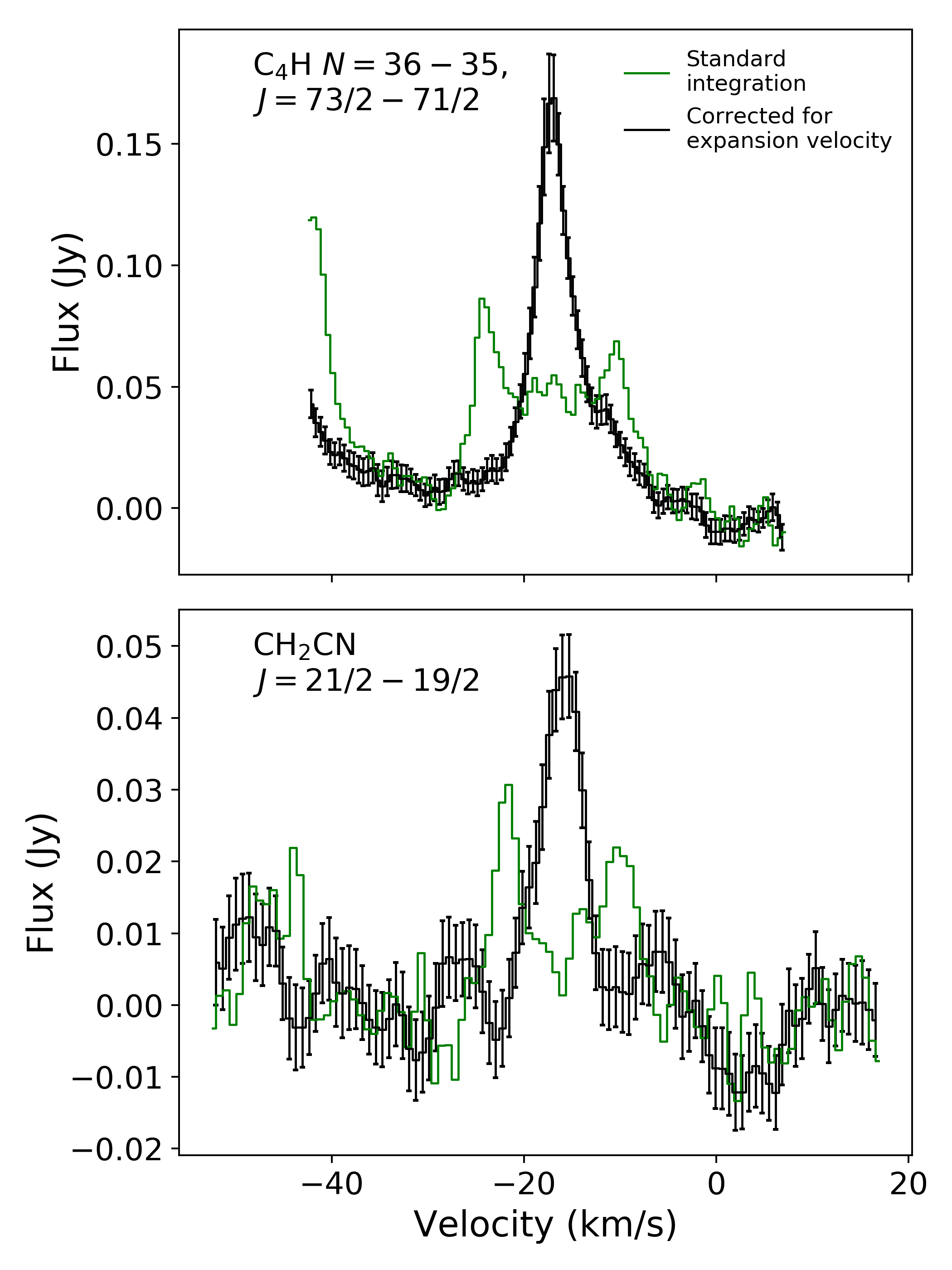}
    \caption{Demonstration of line shifting and stacking algorithm for a strong (top) and weak (bottom) rotational line. Error bars in the corrected spectra denote the standard deviation of flux measurements in a particular velocity bin. Spatial integrations were performed over a deprojected radius range of 0\arcsec -- 1.5\arcsec for \ce{C4H} and 0\arcsec -- 3\arcsec for \ce{CH2CN}.}
    \label{fig:gofish_stack}
\end{figure}

To further investigate these maps within the context of V Hya's circumstellar structure, we utilize the publicly available Python module GoFish \citep{GoFish}. With these tools, we are able to take advantage of the known geometric and velocity properties of the DUDE to shift the spectra associated with each pixel to a common line center and stack them \citep{Yen2016_stackingdisk}, increasing the S/N of detections. Because this software is built for analysis of protoplanetary disks, we modified the Keplerian velocity model it uses to instead account for the expanding nature of V Hya's torus, using the velocity profile presented in \citet{Sahai2022_VHyaDUDE} and the inclination of the disk ($46^{\circ}$). To test the utility this kinematic model has in manipulating spectra at different locations in the DUDE, we performed a velocity-corrected stack for the $N=36-35$ transition of \ce{C4H} and the tentatively-detected the $J=21/2-19/2$ line of \ce{CH2CN} Fig.\ \ref{fig:gofish_stack}. We find here that when the spectra are averaged with a common central velocity, the line profiles become single-peaked, and the signal-to-noise ratios are boosted by over 100\%. This method could thus be a powerful tool in the future for identifying weak molecular signatures in interferometric observations of disk/torus structures surrounding evolved stars, as has proven to be the case for protoplanetary disks \citep[e.g.\ ][]{Matra2017_CO_CO2_fishing}. Refer to Appendix \ref{app:gofish_more} for additional applications of line stacking in these observations.

We also use the velocity-shifted spectra to measure the radial distribution of integrated line intensities in the outflow of V Hya. Using the same physical model of the DUDE, for each transition we compute velocity-integrated line fluxes averaged within annuli at all radii in the disk, and convert the result to brightness temperature units. We restrict the velocity integration of the stacked spectra to $\pm10$\,\kms about the systemic velocity to avoid contributions from base of the high-velocity outflow. It is important to note that this method assumes an infinitely thin 2D structure, and therefore does not take into account the height and flared structure of the DUDE. This means that each bin samples a slightly larger range of radii than expected, since line of sight positions can be contaminated with emission from above/below the mid-plane of the disk (Appendix \ref{app:gofish_more}).

\begin{figure*}
    \centering
    \includegraphics[width=\linewidth]{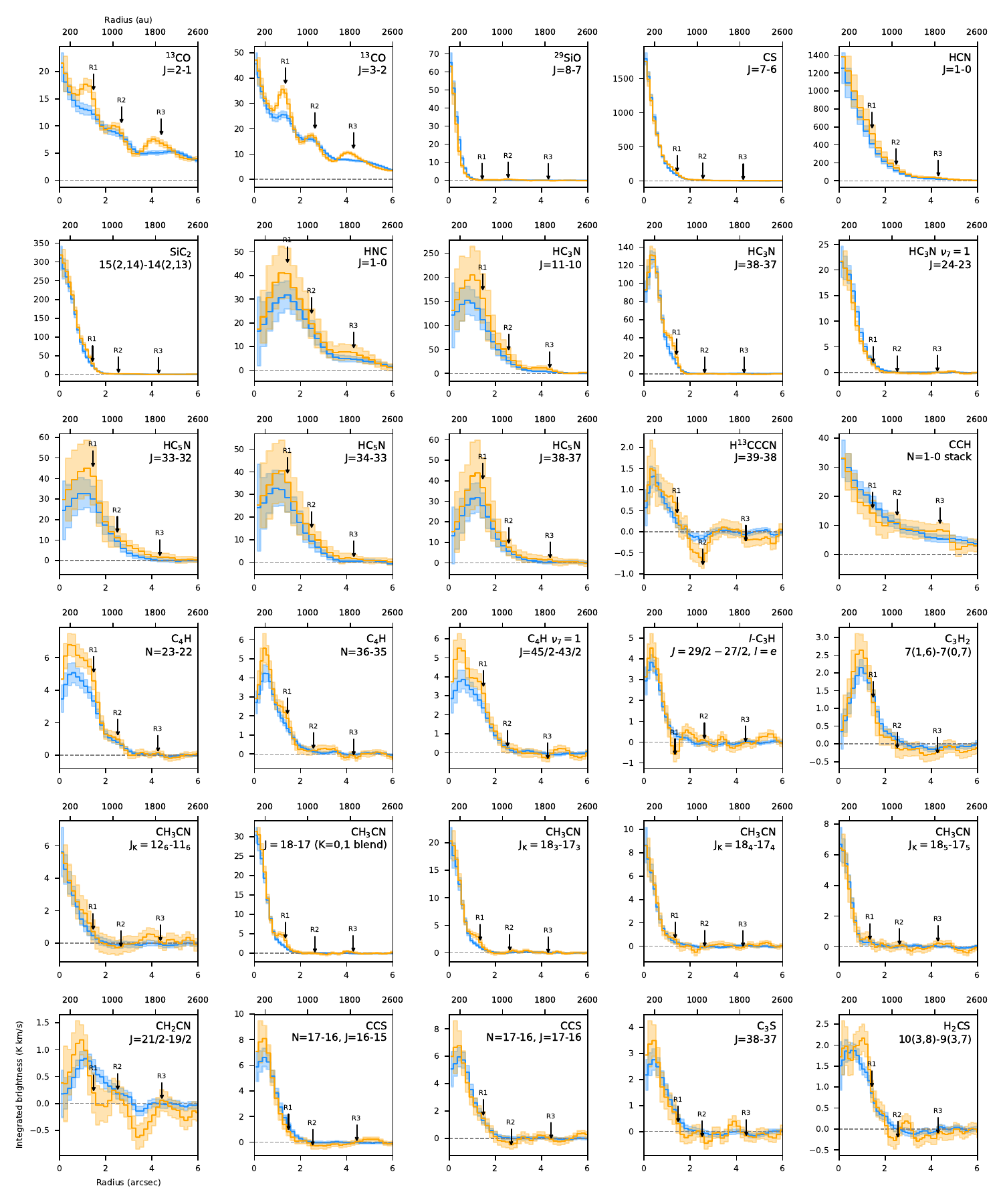}
    \caption{Velocity-integrated radial intensity profiles of unblended, mapped transitions toward V Hya. For each line, two profiles are shown; one corresponding to the average over all angles in the disk (blue), the other extracted over a southward-pointing 100$^{\circ}$ wedge. Shaded regions around both curves denote the standard deviation in each radial bin. Bin sizes are one quarter the major axis of the synthesized beam. Ring structures R1, R2, and R3 are labeled with arrows.}
    \label{fig:rprofs}
\end{figure*}

The deprojected and azimuthally averaged radial profiles for each mapped transition toward V Hya are shown in Fig.\ \ref{fig:rprofs}. The conversion to brightness temperature was done under the Rayleigh-Jeans approximation, and the physical units shown on the upper axis of each plot were calculated using a distance of 400\,pc \citep[eDR3,][]{GAIA2021_dr3}. We do note that \citet{Andriantsaralaza2022_GAIA_AGB} present statistical corrections to GAIA distances for many AGB stars including V Hya (529\,pc) using a Bayesian statistical method based on VLBI measurements of maser sources, as well as new distances calculated using a revised period-luminosity relationship (311\,pc). For consistency with \citet{Sahai2022_VHyaDUDE}, we choose to adopt the uncorrected GAIA distance (400\,pc) in this work.

Looking in general at the intensity profiles in Fig.\ \ref{fig:rprofs}, we find that most transitions have ring-like distributions with a single maximum, occurring at various positions in the disk. We also see that that every transition has a primary maximum at or within the position of R1, and rarely (only in the case of Band 3 lines) do we see any flux outside of R2. The position of R0 can be seen as the maximum for \ce{HC3N} $J=38-37$ and all \ce{C4H} lines; however, there is also evidence of species that peak in intermediate regions of the inner 500\,au, for example, c-\ce{C3H2} and \ce{H2CS} are found slightly inward of R1 but not as compact as R0. In addition to the species with ring-shaped intensity profiles, others show centrally-peaked emission. These transitions can be categorized into two groups: those with flat-topped patterns (e.g.\ \ce{CCS}, \ce{C3S}), and those with more ``cusped" profiles (e.g.\ \ce{CS}, \ce{HCN}). Interestingly, the distributions of \ce{CH3CN} lines appear to modulate from nearly flat-topped to cusped with increasing $K$ values, indicating an energetic dependence on the spatial extent of this molecule. We will explore this effect further in our discussion of excitation and abundances in Section \ref{section:disc}.

Fig.\ \ref{fig:rprofs} also shows a comparison between the global intensity average and that obtained over a 100$^{\circ}$ wedge directed to the south. The latter is included to quantitatively demonstrate spatial asymmetries which tend to occur in this direction. We again see that \ce{C4H} shows consistent anisotropy among its transitions, with its southward profiles appearing ${\sim}$50\% brighter than the average over all angles of the disk. This is the only molecule that shows this degree of angular asymmetry. A similar behavior of southward brightness enhancement can be seen for many lines showing emission at R1 (e.g.\ \ce{HNC}, \ce{CH3CN}, \ce{HC3N}), which is due to a density asymmetry in this ring, as it is apparent in \ce{^{13}CO} as well.

\section{Discussion}
\label{section:disc}
\subsection{Summary of detected species}
\label{section:mol_disc}
The species detected and imaged in this work are important pieces in a broad network of chemical processes at work in the circumbinary waist of V Hya. Here, we will examine this molecular inventory in the context of circumstellar chemistry as a whole, and discuss the observed lines and likely formation mechanisms of each species found in this unique source.

\subsubsection{SiO (silicon monoxide)}
An abundant parent molecule known for its maser emission in M-type AGB envelopes \citep{Gray1999_SiO_masers,Chibueze2016_siomaser_agb}, SiO is the only major carrier of oxygen other than \ce{CO} in C-rich stellar outflows \citep{Agundez_AGB_TE_Review_2020}. Toward V Hya, the $J=8-7$ transition $^{29}$Si isotopologue is detected with a centrally-peaked distribution that is likely optically thick \citep{Sahai2022_VHyaDUDE}. 

\subsubsection{CS (carbon monosulfide)}
\ce{CS} is the primary gas phase reservoir of sulfur in carbon rich stellar winds, and forms out of thermochemical equilibrium in the extended atmospheres of AGB stars \citep{Agundez_AGB_TE_Review_2020}. The emission of this molecule toward V Hya was also discussed in \citet{Sahai2022_VHyaDUDE}, where it is found that the $J=7-6$ transition shown here is optically thick and subthermally excited.

\subsubsection{\ce{SiC2} (silacyclopropynylidene)}
\ce{SiC2} is another important parent molecule in C-rich AGB stars. There is observational evidence that this species is a precursor to SiC grains in the inner regions of stellar envelopes \citep{Massalkhi2018_SiC2_Cstars}, making it important to dust nucleation models. Absorption bands of \ce{SiC2} were detected toward V Hya by \citet{Sarre2000_Cstar_SiC2_abs}, confirming its presence in the upper atmosphere of the carbon star. In these ALMA observations, the $15(2,14)-14(2,13)$ transition is prominent and centrally peaked \citep{Sahai2022_VHyaDUDE}, and we also find evidence of emission lines from the $\nu_3=1$ vibrational state, which lies at $196.37$\,cm$^{-1}$. Though these transitions are too weak to map, their profiles suggest that they arise from central regions of the DUDE where the IR flux is high.

\subsubsection{\ce{HCN} (hydrogen cyanide)}
Along with \ce{C2H2}, \ce{HCN} is one of the most abundant parent molecules in carbon star environments \citep{Agundez_AGB_TE_Review_2020}. In addition to being a major reservoir of carbon, this molecule is important precursor to more complex species, as its photodissociation into  H $+$ \ce{CN} drives the formation of cyanopolyynes and longer carbon chains \citep{Agundez2017_IRC_carbonchains}. In V Hya, we detect the $J=1-0$ transition of this species, which is optically thick due to the high abundance and intrinsic line strength. Like \ce{CS} its average brightness temperature in the central region of the DUDE (${\sim}$60\,K) implies that it may trace warmer regions near the surface of the disk. We also find \ce{HCN} in the high velocity spot. 

\subsubsection{\ce{HNC} (hydrogen isocyanide)}
As a less stable isomer of \ce{HCN}, \ce{HNC} requires additional chemical processing to appear in stellar winds. It is formed primarily through the dissociative recombination of \ce{HCNH+}, which is supplied from cosmic ray and photochemical interactions \citep{Daniel2012_HNC_IRC}. Because of this, HNC typically shows shell-like distributions in carbon-rich outflows. Toward V Hya, we see emission from the $J=1-0$ transition in an large ring peaking at R1 and extending past R3. 

\subsubsection{\ce{HC3N} (cyanoacetylene)}
\ce{HC3N} is a key molecule in kinetic-driven photochemistry surrounding carbon stars. Produced from the reaction between \ce{C2H2} and \ce{CN}, \ce{HC3N} is the first (and most abundant) cyanopolyyne to form, and thus marks the location in an outflow where carbon chain polymerization becomes efficient. \ce{HC3N} was first reported in the circumstellar envelope of V Hya by \citet{Knapp1997_VHya}. In these observations, it is detected at both Band 3 and Band 7, in addition to the $\nu_7=1$ excited state at Band 6. Similar to recent observations of IRC+10216 by \citet{Siebert2022_IRC_HC3N} and \citet{Agundez2017_IRC_carbonchains}, the $J=38-37$ transition shows a more compact emission than the lower energy line $J=11-10$ toward V Hya; however, unlike IRC+10216, these lines do overlap in their spatial distributions toward this source. The $J=24-23$ line of the $\nu_7=1$ state notably shows centrally-peaked emission.

\subsubsection{\ce{HC5N} (cyanodiacetylene)}
Similar to \ce{HC3N}, \ce{HC5N} is primarily formed through diacetylene (\ce{C4H2}) reacting with \ce{CN}. Because this process requires available \ce{C4H2} from hydrocarbon growth, this higher level of complexity is only reached in regions where UV-driven photochemistry is very active \citep[see Fig.\ 5 in][]{Agundez2017_IRC_carbonchains}. In the outflow around V Hya, we see \ce{HC5N} emission peaking near R1; however, interestingly its flux does not drop to zero at the center of the map as it does in other C-rich sources. Though the beam size of ALMA is larger at the $\lambda{\sim}3$\,mm wavelengths where \ce{HC5N} is detected, we do not attribute this behavior to these observational effects, as even in the system velocity channel map where ring emission is at its farthest extent, these transitions are seen at the central position of V Hya. 

\subsubsection{\ce{CCH} (ethynyl radical)}
\ce{CCH} is the main photodissociation product of acetylene (\ce{C2H2}). Because it is quick to undergo neutral-neutral reactions and thereby extend carbon chains, its abundance typically peaks at larger radii than cyanopolyynes, where the extinction of UV photons is lower \citep{Agundez2017_IRC_carbonchains}. Toward V Hya, we find weak emission from a group of features consistent with the spin-coupling hyperfine components of the $N=1-0$ transition of this molecule. The stacked emission of the brightest components ($F=2-1$ and $F=1-0$) reveals a peculiar spatial distribution that is centrally peaked and gradually decreases throughout the DUDE. This is in stark contrast to other AGB envelopes, where \ce{CCH} is almost always found in a shell pattern \citep{Unnikrishnan2024_charting}.

\subsubsection{$l$-\ce{C3H} (propynylidyne radical)}
Next in the sequence of hydrocarbon radicals, linear \ce{C3H} was first detected in IRC+10216 \citep{Thaddeus1985_C3H}, and has been observed in other C-rich AGBs and post-AGBs as well \citep{Unnikrishnan2024_charting,Pardo2007_CRL_fullsurvey}. In our data, we see two lines at Band 7 corresponding to two components of the $J=29/2-27/2$ transition. The distribution is compact, and peaks near R0 like \ce{HC3N}.

\subsubsection{\ce{C4H} (butadiynyl radical)}
\ce{C4H} has two main formation pathways: the first, analogous to \ce{CCH}, is through the photodestruction of \ce{C4H2}; the second is through the reaction \ce{CCH}$+$\ce{C2}. In this sense, this molecule is parallel to \ce{HC5N} in its chain length since it is mainly dependent on the availability of \ce{C4H2}. Because of this, its abundance typically follows a similar distribution to \ce{HC5N} \citep{Cordiner2009_ircmodel_shells}. 

As noted previously, \ce{C4H} is one of the most prominent species found in this study of V Hya, with 16 total transitions detected spanning four different vibrational modes (including the ground state). All observed lines are co-spatial with each other, appearing in a compact ring peaking near R0 (160\,au), with a strong preference toward the southern portion of the DUDE. 

\subsubsection{c-\ce{C3H2} (cyclopropenylidene)}
c-\ce{C3H2} is another important hydrocarbon in C-rich stellar outflows. It is proposed to form in trace amounts as a parent species, but in regions of kinetic-driven chemistry c-\ce{C3H2} can be synthesized from barrierless reactions involving \ce{CH}, \ce{CCH}, and \ce{C3H} \citep{VandeSande2018}. This species was shown to be coincident with \ce{C4H} toward the carbon star IRAS 15194-5115 by \citet{Unnikrishnan2024_charting}, and its observed excitation temperature of 20\,K in IRC+10216 also implies a shell-like distribution \citep{Kawaguchi1995_IRCsurvey}. Toward V Hya, the emission we observe from the $7(1,6)-7(0,7)$ transition shows a hollow distribution with a maximum at ${\sim}400$\,au. This is notably at a larger radius than \ce{C4H}, which may result from the smaller upper state energy of the c-\ce{C3H2} transition.

\subsubsection{\ce{CH3CN} (methyl cyanide)}
Methyl cyanide is one of the more complex species we detect toward V Hya, in addition to being a useful temperature probe due its symmetric top geometry. This molecule is produced mainly through the dissociative recombination of \ce{CH3CNH+}, an intemediate formed from the radiative association of \ce{CH3+} and \ce{HCN}. Despite this being a process predominantly driven by photochemistry (in addition to cosmic rays), \citet{Agundez2015_CH3CNmaps_IRC+10216} report a very compact (though still hollow) abundance distribution of this molecule in the envelope of IRC+10216 which could not be reproduced by chemical models. We detect the $J=18-17$ family of lines toward V Hya, in addition to the $K=6$ component of $J=12-11$. These transitions show centrally-peaked integrated intensity maps, though inspection of the systemic velocity channel map (Fig.\ \ref{fig:chmaps}) shows primary peaks located symmetrically at $r{\sim}$0.35\arcsec from V Hya.

\subsubsection{\ce{CH2CN} (cyanomethyl radical)}
To date, \ce{CH2CN} has only been observed in the envelope of IRC+10216, where its column density is less than \ce{CH3CN} by a factor of three \citep{Agundez2008_newIRCmols}. This species is formed from the same dissociative recombination that is responsible for \ce{CH3CN}, with a comparable branching ratio between the two products \citep{Loison2014_HCN_HNC}. We initially label \ce{CH2CN} as a tentative detection toward V Hya because only one transition ($J=21/2-19/2$) is observed, and it is quite weak; however, as discussed in Section \ref{section:profs}, applying a pixel-by-pixel velocity correction increases this significance notably. The intensity map and radial profile are consequently low S/N, though emission is seen slightly interior to R1, similar to c-\ce{C3H2}.

\subsubsection{\ce{CCS} (dicarbon sulfide)}
Analogous to the polyyne and cyanopolyyne chemistry discussed above, sulfur-terminated carbon chains are also known to form in AGB environments. In the ISM, a common pathway to \ce{CCS} is through the exothermic reaction between atomic S and \ce{CCH} \citep{Petrie1996_CCS_C3S}, which is likely dominant in evolved stars due to the availability of acetylene and \ce{CCH}. CCS is known in IRC+10216, as well as three additional carbon stars studied in \citet{Unnikrishnan2024_charting}. Here, we detect two transitions of \ce{CCS} toward V Hya. The integrated maps show intensity peaks near R0 with a relatively flat profile inward. This distribution is likely very similar to \ce{HC3N} $J=38-37$, though appears less ring-like due to the larger beam size at Band 6 where \ce{CCS} is detected.

\subsubsection{\ce{C3S} (tricarbon monosulfide)}
\ce{C3S} can be formed through direct C addition onto \ce{CCS} or through another neutral-radical reaction between \ce{CCH} and \ce{CS} \citep{Millar2001_CnS_chains}. We report one cleanly detected transition of this molecule toward V Hya at 219\,620\,MHz, which shows a similar distribution to \ce{CCS}, though with a slight enhancement to the southern portion of the DUDE.

\subsubsection{\ce{H2CS} (thioformaldehyde)}
\ce{H2CS} is bserved toward IRC+10216, and the proto-planetary nebula CRL 2688 \citep{Agundez2008_newIRCmols,Zhang2013_CRL2688survey}. The main chemical route to this species involves a neutral--neutral reaction between \ce{CH3} and atomic \ce{S} \citep{Woods2003_CRL618model}. In V Hya, we detect blended emission from the $10(3,8)-9(3,7)$ and $10(3,7)-9(3,6)$ transitions, as well as a very weak feature that could be attributed to the $10(4,7)-9(4,6)$ line. The stacked intensity map of the former two lines shows flux out to R1, and a peculiar east-west asymmetry that is not seen from any other molecule in V Hya.

\subsection{Excitation conditions of detected species}
\label{section:methyl_pops}

With some species detected toward V Hya with multiple rotational lines, a population study is useful in constraining excitation conditions in its circumstellar environment. Here, we employ two methods to measure rotational temperatures of molecules in our sample.For species with more than one unblended transition detected (in our case \ce{^{13}CO}, \ce{C4H}, \ce{SiC2}), we use a radial population diagram \citep[similar to the methods in][]{Loomis2018_rotationdiag}. Following the formalism of \citet{Goldsmith1999_populationdiagram}, the Boltzmann equation for rotational populations can be arranged in the following way:

\begin{equation}
\label{eq:popdiag}
    \ln{\frac{N_u}{g_u}}=\ln{\left(\frac{N_T}{Q\left(T_{\mathrm{rot}}\right)}\right)}-\frac{E_u}{kT_{\mathrm{rot}}}
\end{equation}

\begin{figure*}[t!]
    \centering
    \includegraphics[width=\linewidth,trim=1.5cm 0.5cm 1.9cm 1cm,clip]{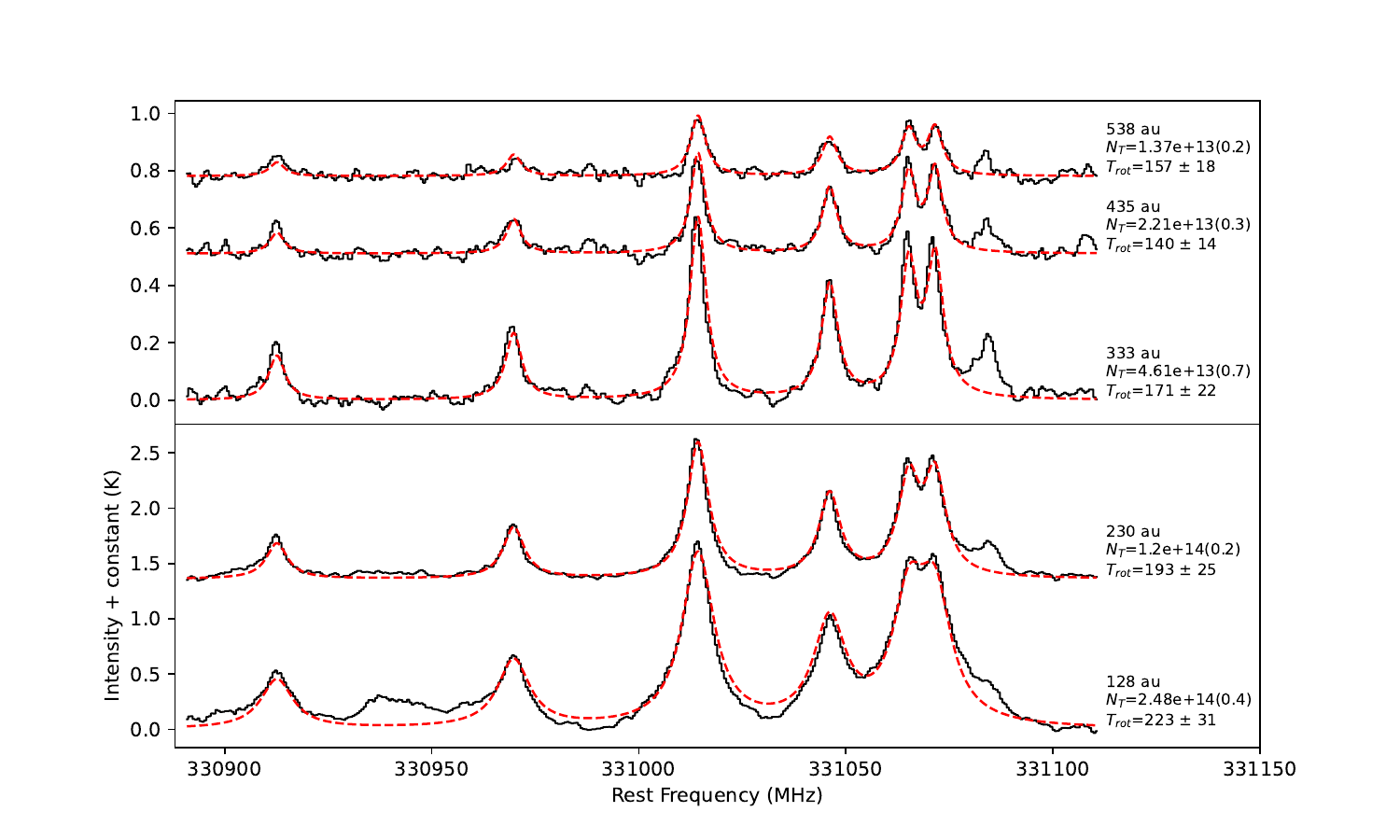}
    \caption{LTE spectral fit (dashed red) to observed velocity-stacked \ce{CH3CN} $J=18-17$ group of transitions at several radii (black) in the disk of V Hya. From left to right, the detected peaks are $K=5,4,3,2,1,0$. Spectra were extracted at increments of one half the ALMA beam width, and molecular parameters were obtained from CDMS \citep{2005JMoSt.742..215M}.}
    \label{fig:CH3CN_fits}
\end{figure*}

Where $N_u$ is the column density of the upper state, $g_u$ is the degeneracy of the upper state, $N_T$ is the total column density, $Q$ is the partition function, $E_u$ is the upper state energy, and $T_{\mathrm{rot}}$ is the rotational excitation temperature. We obtain $\frac{N_u}{g_u}$ using the optically-thin intensity approximation:

\begin{equation}
\label{eq:nthin}
    \frac{N^{\mathrm{thin}}_u}{g_u}=\frac{3k\int T_{mb}dv}{8\pi^3\nu S_{ij}\mu^2}
\end{equation}

where $\nu$ is the transition frequency, $S_{ij}\mu^2$ is the line strength, and $\int T_{mb}dv$ is the velocity-integrated main beam brightness temperature. This allows us to plot $\ln\frac{N_u}{g_u}$ as a function of $E_u$ and obtain $N_T$ and $T_{\mathrm{rot}}$ from a linear fit to Eq.\ \ref{eq:popdiag}. This method is then performed on each radial measurement of $\int T_{mb}dv$  \footnote{In the case of \ce{^{13}CO}, the value of $T_{mb}$ at the system velocity is used instead of the integrated intensity, due to a nearby blend between the $J=2-1$ line and \ce{H^{13}CCCN} $J=25-24$. While this does not affect the slope of the rotation diagram, the column density is consequently not available}. For each molecule, the ALMA images were smoothed to a common resolution before velocity-shifting, matching that of the lowest frequency line. The results of this method can be seen in Fig.\ \ref{fig:pop_diag_twomol}, and corresponding LTE simulations under the derived parameters can be found in Appendix \ref{app:all_fits}. Both \ce{SiC2} and \ce{C4H} show a clear decreasing trend in rotational temperature with radius. While the rotation diagrams here are useful in obtaining a first glance at their excitation conditions, they are limited by the very small number of transitions constraining the populations. Because the Boltzmann distributions are not well-sampled, additional transitions could have a large effect on the calculated temperature.In contrast, \ce{CH3CN} is detected with five observed lines in the $J=18-17$ family spanning a wide range of energy states, making it the best temperature probe among this sample. However, due to the nearby spacing of these $K$-components, their emission overlaps, especially at small radii where the line width is larger (see Appendix \ref{app:gofish_more}). Therefore, a rotation diagram using direct-integrated $\int Tdv$ measurements is not straightforward, so for this molecule we instead utilize a direct LTE fit to the radially-extracted, velocity-stacked observations of this family of transitions. The fitting procedure assumes optically thin emission, and an empirically measured line profile (Appendix \ref{app:gofish_more}), and minimizes the least-squares difference between simulation and model in regions of the spectrum within 15\,\kms of a $K$-component. Figure \ref{fig:CH3CN_fits} shows the best-fit spectra.

\begin{figure}[t!]
    \centering
    \includegraphics[width=\linewidth]{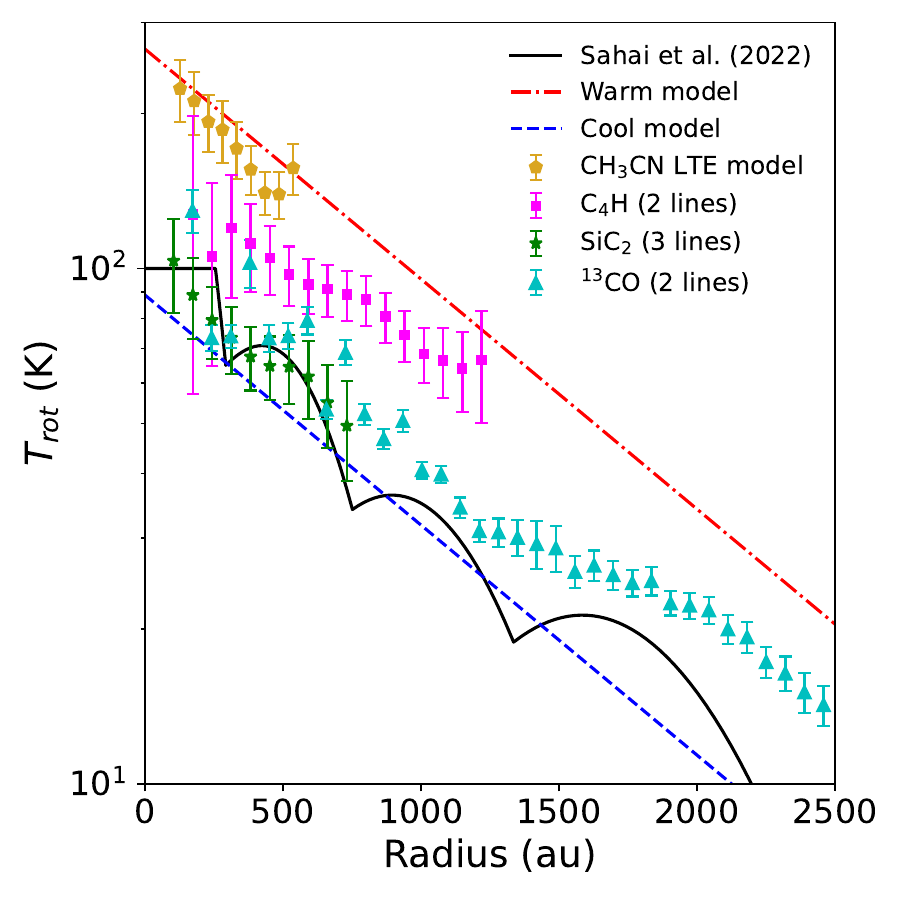}
    \caption{Measured rotational temperatures as a function of radius in the circumstellar disk of V Hya. For most species (\ce{C4H}, \ce{SiC2}, \ce{^{13}CO}) a two-point population ratio or rotation diagram was used, while \ce{CH3CN} was fit with a full LTE spectral model due to blends between $K$-components (see text). Solid line denotes the temperature profile found in \citet{Sahai2022_VHyaDUDE}. Empirical bounds on the excitation temperatures derived from these species are shown in red and blue dashed lines.}
    \label{fig:pop_diags}
\end{figure}

The results of all rotational temperature measurements are shown together in Fig.\ \ref{fig:pop_diags}, along with the temperature profile derived by \citet{Sahai2022_VHyaDUDE} using the brightness temperature of optically thick \ce{^{12}CO}. We find that the excitation temperatures vary by up to a factor of ${\sim}2$, with \ce{SiC2} and \ce{^{13}CO} reaching a maximum of 100\,K in the inner DUDE, and \ce{CH3CN}${\sim}200$\,K. The decreasing trend in all derived temperatures matches the global log-linear slope of the \ce{^{12}CO} surface brightness; however, we find no variation of the temperature around the ring structures R1, R2, R3, as was found with that method.

\begin{figure}[t!]
    \centering
    \includegraphics[width=\linewidth,trim=0.3cm 0.1cm 0.2cm 0.2cm,clip]{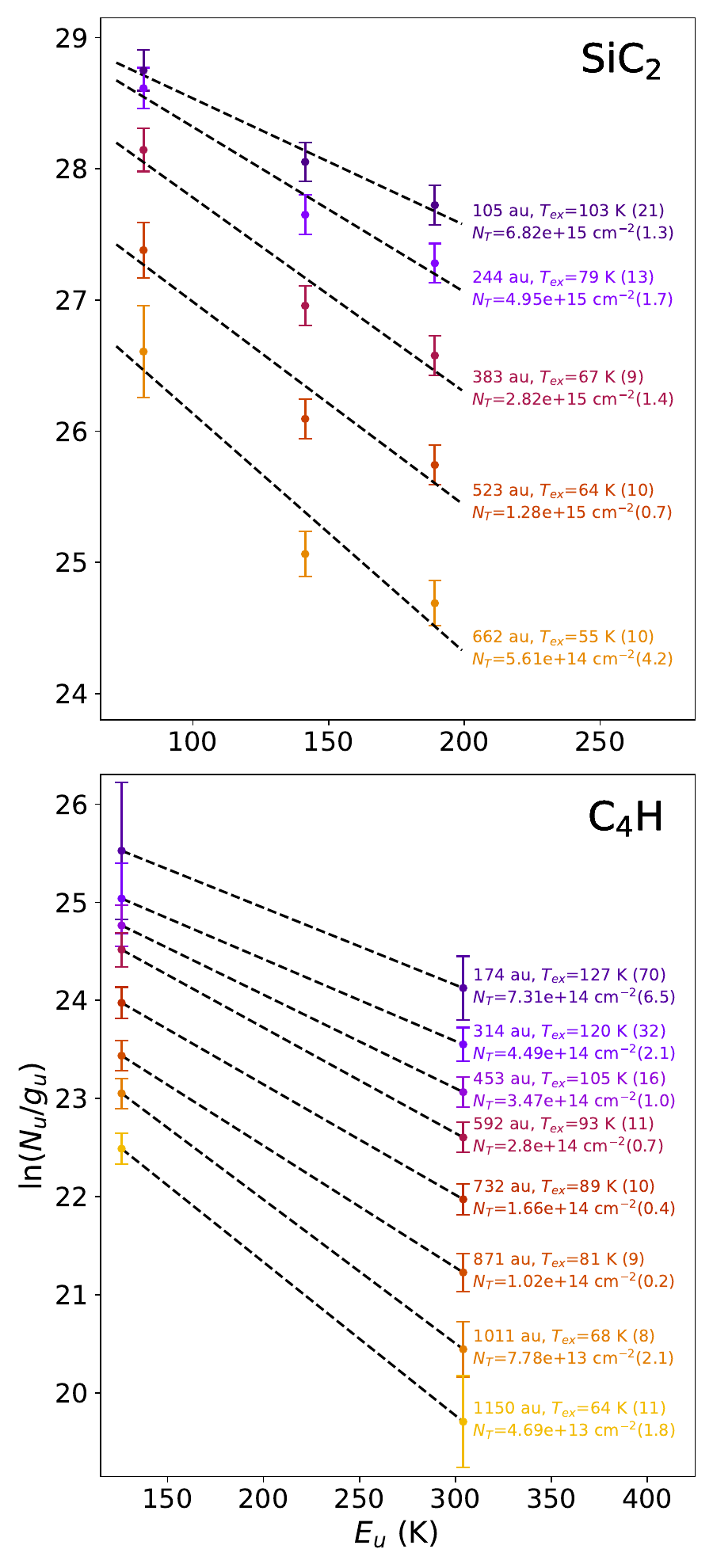}
    \caption{Rotation diagrams from ALMA-observed transitions of \ce{SiC2} (top) and \ce{C4H} (bottom) toward V Hya. The lines (left to right) are $J_{K_a,K_c}=11_{2,10}-11_{0,11}$, $15_{2,14}-14_{2,13}$, and $14_{6,9}-13_{6,8}$ for \ce{SiC2}; and $N=23-22$ and $N=36-35$ for \ce{C4H}. For each radius, the calculated excitation temperature and column density is listed. Radial bin sizes shown here are half the ALMA beam.}
    \label{fig:pop_diag_twomol}
\end{figure}

If local thermodynamic equilibrium (LTE) is assumed, we find that the kinetic temperatures implied by \ce{CH3CN} and \ce{C4H} are considerably larger than those derived in \citet{Sahai2022_VHyaDUDE}. One possible explanation for this discrepancy could be  that emission from these species is arising from warmer vertical regions of the DUDE than \ce{^{12}CO}. Another more likely scenario is that we are observing an effect of infrared (IR) pumping, where molecules are radiatively excited to higher vibrational states, and subsequently return to the ground state with larger rotational energy. This behavior was first observed for \ce{CH3CN} in the Orion molecular cloud by \citet{Goldsmith1983_ch3cn_hc3n_vib}, and \citet{Agundez2008_newIRCmols} note that it can produce superthermal excitation temperatures for certain molecules in the envelope of IRC+10216. \citet{Massalkhi2019_CS_SiO_SiS_Cstars} showed that IR-pumping can yield $T_{\mathrm{ex}}/T_{\mathrm{kin}}>4$ for transitions of CS, SiO, and SiS in other carbon stars, which is consistent with the discrepancy in temperature laws shown in Fig.\ \ref{fig:pop_diags}. Though we did not detect any transitions from vibrationally excited states of \ce{CH3CN}, the number of these lines we see from other species like \ce{C4H} and \ce{HC3N} suggests that IR pumping is an important excitation mechanism in V Hya. 

An key distinction between the result here and that observed in \citet{Goldsmith1983_ch3cn_hc3n_vib} is that the relative line strengths of ground state \ce{CH3CN} are well-characterized by a single excitation temperature, as opposed to having $T_{\mathrm{rot}}$ vary among K-states. This suggests that the rotational populations have still equilibrated to a Boltzmann distribution through a mix of collisional and radiative processes. Therefore, although the excitation temperatures may not reflect the kinetic temperature in the DUDE, they can still be used it as an estimate to derive the column densities and abundances using Eq.\ \ref{eq:popdiag}. 

Based on the results of these population analyses, we define two empirical temperature laws (``warm" and ``cool" models in Fig.\ \ref{fig:pop_diags}) which bound the measured excitation characteristics in V Hya. Both temperature models are exponentials with
\begin{equation}
    T(r) = T_1\alpha^{\left(1-r/r_1\right)}
\end{equation}
where $T_1$ is the temperature at the R1 ring (560\,au), and $\alpha$ is the slope. The warm and cool models differ by a factor of three in magnitude ($T_{1,\mathrm{cool}}=50$\,K and $T_{1,\mathrm{warm}}=150$\,K), and both have $\alpha=1.78$. Given the variance in measured excitation temperatures, and that \ce{CH3CN} often displays the highest excitation temperatures in these environments \citep{Agundez2008_newIRCmols}, we expect that other molecules will fall between these two curves. So for species with limited line detections and thus no available rotational temperatures, these temperature profiles can be used as conservative limits on their excitation conditions.

\subsection{Abundance distributions}

\begin{figure*}[t!]
    \centering
    \includegraphics[width=\linewidth,trim=3.5cm 1.5cm 3.5cm 1cm,clip]{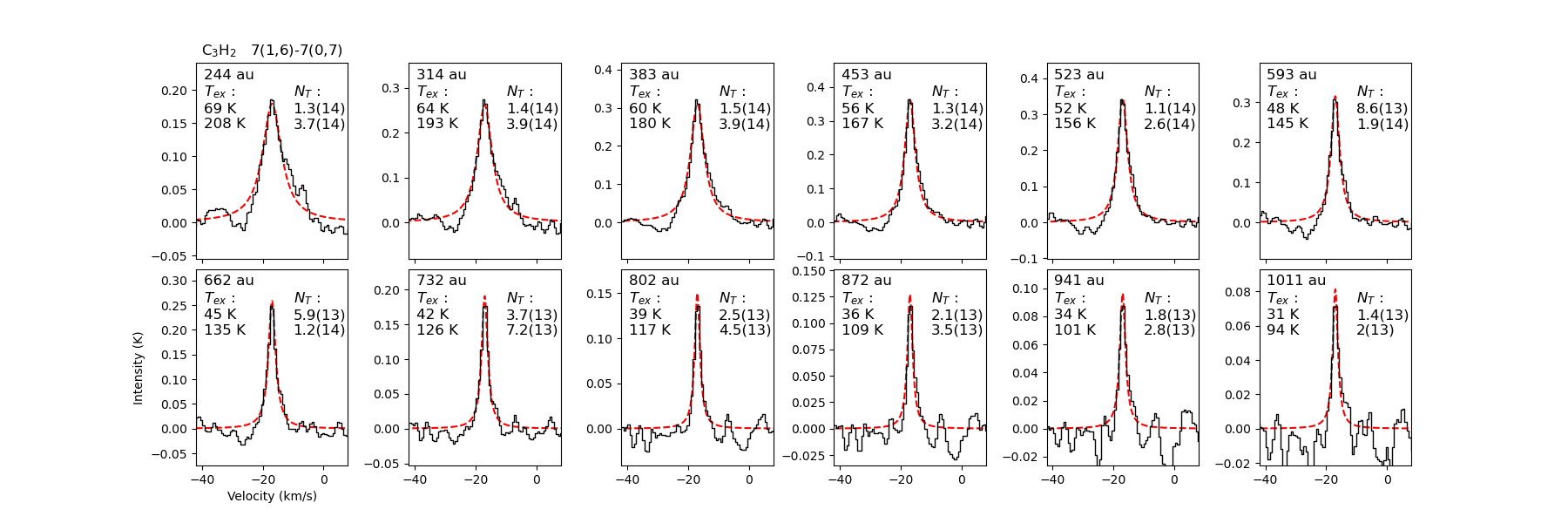}
    \caption{Single-line LTE model of the $J_{K_a,K_c}=7_{1,6}-7_{0,7}$ transition \ce{c-C3H2} emission (red) at radii where it is detected toward V Hya. Black line shows velocity-shifted and stacked ALMA observations. On each panel, two retrieved column densities for the molecule are listed in units cm$^{-2}$, with the exponent written in parentheses. The two column densities correspond to the best-fit LTE model when assuming the cool (top value) and warm (bottom) excitation temperature models of the DUDE. The temperatures of these models are shown on the left side of each spectrum. Only one simulation is shown in each panel, as the best-fit LTE spectrum is the same for both temperature assumptions.}
    \label{fig:fit_c3h2}
\end{figure*}

\begin{figure*}[t!]
    \centering
    \includegraphics[width=0.9\linewidth,trim=0.5cm 0.5cm 0.3cm 0.1cm,clip]{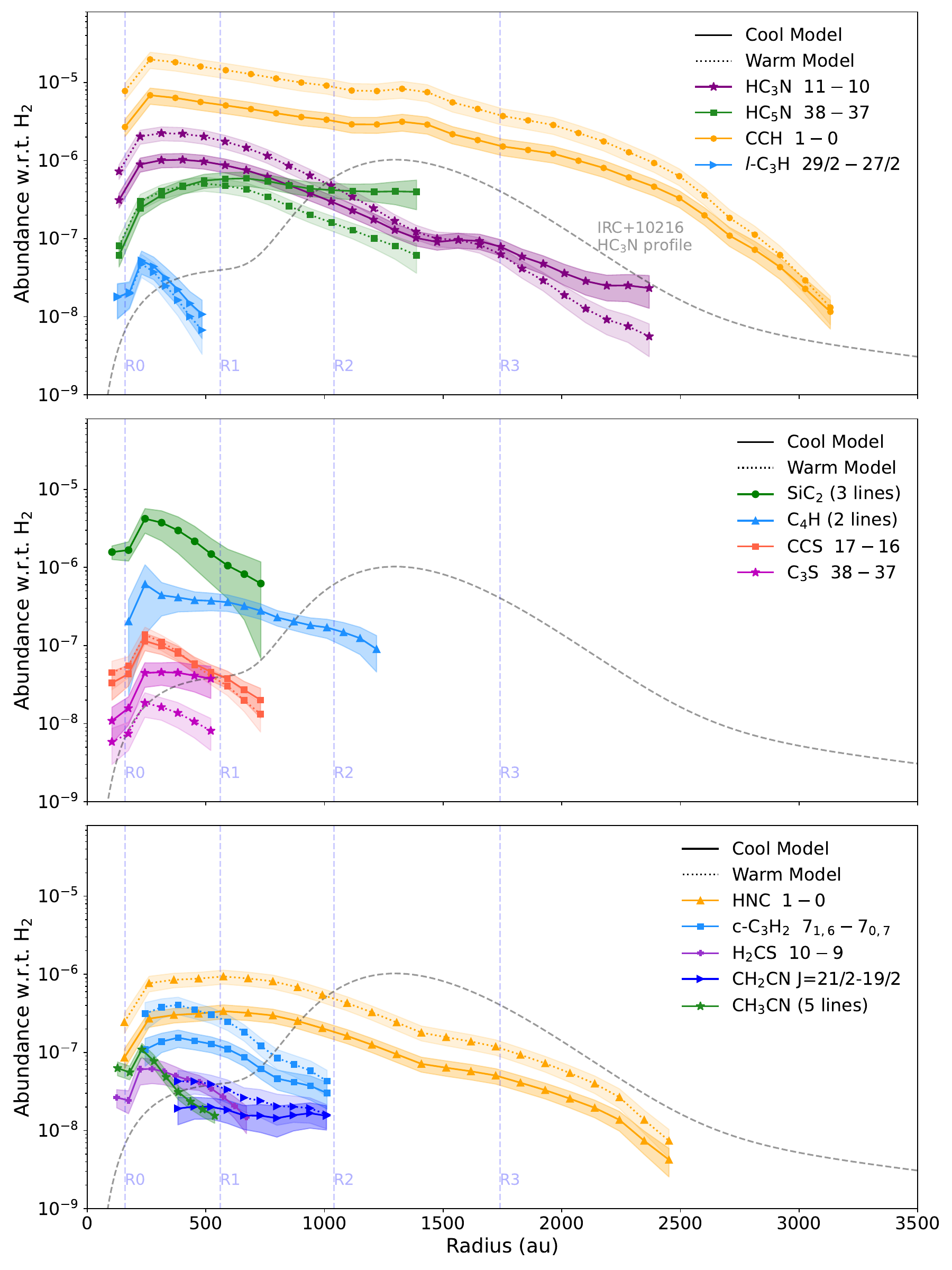}
    \caption{LTE-calculated radial abundance profiles of molecules detected toward V Hya. For species with no measured excitation temperature, two abundances are shown, corresponding to the cool (solid) and warm (dotted) temperature bounds discussed in Section \ref{section:methyl_pops}. The shaded region around each profile denotes the uncertainty, including the standard deviation of flux in the radial bin, the noise level of the integrated image, and the ALMA flux calibration error. Radial spacing of points is one-quarter of the beam size, and values are only shown where the azimuthally averaged line flux is larger than 3 times the rms of the image. The drop in all abundances near the center of the outflow is caused by a factor of ${\sim}2.5$ depletion of \ce{CO} in this region measured by \citep{Sahai2022_VHyaDUDE}.}
    \label{fig:LTE_abunds}
\end{figure*}

\subsubsection{LTE abundance estimates}
With new information on molecular excitation and temperature structure in the DUDE, we now aim to place estimates on the abundances of various species detected toward V Hya. Here, we will focus mainly on daughter species, as they are important in constraining circumstellar chemistry, and have more extended distributions resulting in lower optical depths.

As discussed previously, rotational populations in the DUDE are a result of a complex combination of collisional and radiative pumping processes. Unfortunately, a full simulation of these mechanisms for every molecule is not feasible due to a lack of an IR flux model throughout the DUDE, and unknown collisional excitation rates and vibrational band strengths for many of the species of interest. Therefore, in order to calculate abundances, we first employ an LTE approach adopting excitation models derived in the previous section. In all other cases, we perform two abundance calculations. This casts a wide range in physical conditions, allowing us to still place strict bounds (within an order of magnitude) on abundances despite lacking a detailed excitation model.

To calculate the abundance profiles, we first determine the radial column densities for each species. For molecules with excitation temperatures (\ce{CH3CN}, \ce{C4H}, \ce{SiC2}), the column density is available from directly from the population diagram or spectral fit (\ce{CH3CN}). For all other species (mostly detected with only one line), we use a single-line least-squares LTE fit to their velocity-corrected spectra\footnote{One exception to this is \ce{H2CS}, which has a group of lines detected around 343.4\,GHz. A temperature fit is not possible due to weak line strengths, and blends with \ce{C4H} $\nu_7=1$, so in this case the LTE analysis was performed assuming only the cool temperature model (Fig.\ \ref{fig:fit_h2cs}), which provided better fits to the full group than the warm model.}. We perform this fit twice assuming fixed excitation temperatures corresponding to the cool and warm temperature models discussed in Section \ref{section:methyl_pops}. An example of the resulting radial model spectra is shown with \ce{c-C3H2} in Fig.\ \ref{fig:fit_c3h2}. A Lorentzian line shape is used, as it fits the stacked line shapes very well, especially for large radii. Below a radius of ${\sim}$100\,au, line profiles are much wider and very irregular, so we restrict our analysis to radii larger than this threshold as they result in the best fits. More details on the fitting procedure, as well as all best-fit spectra, are presented in Appendix \ref{app:gofish_more}.

After obtaining column densities for the molecules of interest, we use the integrated brightness of \ce{^{13}CO} $J=3-2$ along with Eq.\ \ref{eq:nthin} to obtain the radial column density of \ce{^{13}CO}, and calculate the molecular abundance relative to \ce{H2} with the following:

\begin{equation}
\label{eq:abund}
    f\left(r\right) = \frac{N_T\left(r\right)}{N_{\mathrm{\ce{^{13}CO}}}\left(r\right)}\times\frac{\left[\mathrm{\ce{^{13}C}}\right]}{\left[\mathrm{\ce{^{12}C}}\right]}\times f_{^{12}\mathrm{CO}}
\end{equation}
using a \ce{^{12}C}/\ce{^{13}C} ratio of 70 \citep{Sahai2022_VHyaDUDE}. In \citet{Sahai2022_VHyaDUDE}, it was found that the abundance of \ce{^{12}CO} is $10^{-3}$ throughout most of the DUDE; however, in the central regions this value drops by at least a factor of ${\sim}$ 2.5. So for radii less than 200\,au, $f_{^{12}\mathrm{CO}}$ is instead taken as $4\times10^{-4}$.

To ensure that the numerator and denominator in Eq.\ \ref{eq:abund} are sampled on the same spatial scales, we convolve the \ce{^{13}CO} $J=3-2$ map to match the beam size for each respective molecule, and recalculate the \ce{^{13}CO} brightness profile with the new resolution. Abundances are then derived for all radii in the DUDE where the integrated line brightness is larger than 3 times the rms, with radial bin spacings at one-quarter the major axis of the beam.

The results of this analysis are shown in Fig.\ \ref{fig:LTE_abunds}. Because the large mass-loss rate of V Hya \citep[$>$10$^{-5}$\,\msunyr; ][]{Knapp1997_VHya} is similar to that of IRC+10216, we also show the measured abundance profile of \ce{HC3N} in that source \citep{Siebert2022_IRC_HC3N} as a reference for the typical distribution of a daughter molecule in C-rich AGBs. In general, the observed abundance profiles follow nearly log-linear or single-peaked distributions with a central minimum due to the drop in the \ce{CO} abundance in this region, and the radii where these molecules are most abundant are all within 500\,au. Furthermore, we see no correlation between the trends here and the locations of the density enhancements R0--R2, meaning the chemistry appears to be independent of these physical structures. In Table \ref{table:av_abunds}, we provide the average abundances and column densities of each molecule in V Hya under the two temperature assumptions, along with their observed abundances in IRC+10216. The difference between results obtained from the two adopted temperature profiles is typically less than a factor of two. The exceptions to this occur in cooler regions of the DUDE (outside R2) for the carbon chains \ce{HC3N} and \ce{HC5N}.

The observed abundances and column densities of the cyanopolyynes are in good agreement with those seen in the prototypical source IRC+10216 and C-rich AGBs with similar mass-loss rates \citep[II Lup, V358 Lup][]{Unnikrishnan2024_charting}, with the abundance of \ce{HC3N} reaching a maximum of ${\sim}10^{-6}$ and \ce{HC5N} showing slightly lower density and peaking farther out in the wind. However, the distribution of these carbon chains is unique to this source, as the observed peak of \ce{HC3N} abundance in V Hya (400\,au) is found to be more than a factor of three smaller than what is seen for IRC+10216 \citep[1300\,au; ][]{Agundez2017_IRC_carbonchains}. Since \ce{HC3N} is regarded as the first photochemistry product to arise in carbon star envelopes, the observation of \textit{all} of carbon chains (\ce{HC3N}, \ce{HC5N}, \ce{CCH}, \ce{C4H}) peaking within $r{\sim500}$\,au of V Hya is evidence that their production is markedly different from IRC+10216, considering its very similar mass-loss rate.

The obtained abundance for \ce{C4H} is notably smaller in V Hya (1--3$\times10^{-7}$) than in other C-rich AGBs, where it is normally similar to that of \ce{HC3N}. A likely explanation for this is that an increased population of \ce{C4H} is present in vibrationally excited states, from which we have detected numerous rotational transitions (Table 2). If we take the $J=45/2-43/2$ transition of the $\nu_7=1$ state as an example and apply Eq.\ \ref{eq:nthin} with an excitation temperature of $150$\,K (assuming the same \ce{H2} column density as the ground vibrational state), we find an estimate fractional abundance of $4\times10^{-7}$ for \ce{C4H}\,${\nu_7=1}$ in V Hya, which is comparable to the values obtained for the ground state. Thus, given that \ce{C4H} is present in much warmer regions around V Hya than in spherical AGBs like IRC+10216, IR excitation to its many vibrational states is likely much more efficient in this object, causing the observed underabundance of ground state \ce{C4H} relative to those sources.

The distribution of \ce{C4H} density shown in Fig.\ \ref{fig:LTE_abunds} is also found to be relatively flat or increases steadily in the regions sampled, unlike \ce{HC3N} and \ce{HC5N} where we have found a clear growth \textit{and} decrease of the abundance. The average abundance of $l$-\ce{C3H} is two orders of magnitude less than \ce{CCH}, consistent with this observed ratio in IRC+10216 \citep{Thaddeus1985_C3H}.

The large abundance of \ce{CCH} (2--8$\times10^{-6}$) also matches observations of IRC+10216; however, we find no evidence of a decrease towards inner radii other than the drop due to the \ce{CO} abundance. This is in stark contrast to other carbon star envelopes where \ce{CCH} is always found in a shell-like distribution at a radius peaking outside that of \ce{HC3N} \citep{Unnikrishnan2024_charting,Agundez2017_IRC_carbonchains}. Since our observations do not sample radii less than ${\sim}100$\,au, it is possible that the \ce{CCH} abundance has a steep drop-off within this distance that we are not able to detect. However, from the available information, it appears that this photodissociation product is available immediately in the circumbinary environment of V Hya.

The sulfur-terminated carbon chains show another slight discrepancy from their presence in IRC+10216. The average abundance ratio of $\left[\mathrm{\ce{C3S}}\right]/\left[\mathrm{\ce{CCS}}\right]$ calculated here is ${\sim}0.2$ in V Hya, whereas in IRC+10216 this ratio is closer to unity \citep{Bell1993_C3S_C5S}. This could be owing to non-LTE effects, as the $J=38-37$ of \ce{C3S} may be sub-thermally excited resulting in underestimated abundances; however, an exact analysis of this is not presently possible due to a lack of collisional rates for this molecule. Using \ce{HC3N} as a proxy, the critical density of the observed line would be $5\times10^{6}$\,cm$^3$ \citep{Siebert2022_IRC_HC3N} scaled by the ratio of Einstein $A_{ij}$ values of these transitions for \ce{C3S} ($\log A_{ij}=-3.08$) and \ce{HC3N} ($\log A_{ij}=-2.48$), respectively. This gives a value of ${\sim}1\times10^{6}$\,cm$^3$, which is larger than the average gas density within R1 ($10^5$\,\cmnegthree), meaning the transition could be subthermally excited.

The measured eak abundance of \ce{CH3CN} (${\sim}10^{-7}$) is consistent with its peak abundance in IRC+10216 \citep{Agundez2015_CH3CNmaps_IRC+10216}, but like \ce{CCH} its abundance appears to be strictly increasing inward, until the global drop in all abundances at 200\,au due to \ce{^{12}CO}. The abundance ratio of $\left[\mathrm{\ce{CH3CN}}\right]/\left[\mathrm{\ce{H2CS}}\right]{\sim}1$ is slightly higher than observed toward IRC+10216 \citep{Agundez2008_newIRCmols}, and both \ce{H2CS} and \ce{CH2CN} appear to have a more extended distributions in V Hya (with abundances $>10^{-8}$ between R1 and R2). c-\ce{C3H2} shows a clear ring distribution peaking around ${\sim}400$\,au with an abundance matching those found for the carbon stars in studied in \citet{Unnikrishnan2024_charting}.

We stress that the results in Fig.\ \ref{fig:LTE_abunds} and Table \ref{table:av_abunds} are initial estimates for this source, as they are mostly derived from one or two lines of each molecule, assuming temperature profiles from this work and \citet{Sahai2022_VHyaDUDE}. To reduce the uncertainty to within a factor of two and confirm peculiarities we highlight for daughter species in the disk of V Hya, future wideband surveys are necessary as they will offer crucial information on the full ro-vibrational populations of these chemical products. 

\subsubsection{Radiative transfer calculations}
\label{section:radex}

\begin{deluxetable*}{lccccccccc}[t!]
    \tablecaption{Average column densities and abundances of daughter molecules in the disk of V Hya}
    \tablewidth{\columnwidth}
    \tablehead{
    \colhead{Molecule} & \colhead{$r_{\mathrm{min}}$\,[au]}& \colhead{$r_{\mathrm{max}}$\,[au]} & \colhead{$N_1$\,[\cmnegtwo]} & \colhead{$N_2$\,[\cmnegtwo]} & \colhead{$f_{\mathrm{LTE}1}$} & \colhead{$f_{\mathrm{LTE}2}$} & \colhead{$f_{\mathrm{RADEX}}$} & \colhead{$f_{\mathrm{IRC+10216}}$} & \colhead{ref} } 
    \startdata
SiC$_2$ & 105 & 732 & $2.9\left(15\right)$ & -- & $2\left(-6\right)$ & -- & -- & $2\left(-7\right)$ & 1\\
HC$_3$N & 134 & 2369 & $2.9\left(14\right)$ & $6.1\left(14\right)$ & $3.4\left(-7\right)$ & $6.5\left(-7\right)$ & $2.3\left(-7\right)$ & $1.4\left(-6\right)$ & 2\\
HC$_5$N & 158 & 1529 & $2.6\left(14\right)$ & $2.8\left(14\right)$ & $3.8\left(-7\right)$ & $3.3\left(-7\right)$ & -- & $2\left(-7\right)$ & 2\\
CCH & 159 & 3132 & $1.7\left(15\right)$ & $4.7\left(15\right)$ & $2.3\left(-6\right)$ & $6.1\left(-6\right)$ & $1.9\left(-6\right)$ & $3\left(-6\right)$ & 2\\
$l$-C$_3$H & 127 & 484 & $3.9\left(13\right)$ & $3.5\left(13\right)$ & $2.7\left(-8\right)$ & $2.3\left(-8\right)$ & -- & $5\left(-8\right)$ & 2\\
C$_4$H & 174 & 1220 & $2.6\left(14\right)$ & -- & $2.8\left(-7\right)$ & -- & -- & $1.1\left(-6\right)$ & 3\\
HNC & 157 & 2452 & $1\left(14\right)$ & $2.8\left(14\right)$ & $1.4\left(-7\right)$ & $3.7\left(-7\right)$ & $8.1\left(-8\right)$ & $5.8\left(-8\right)$ & 4\\
CCS & 104 & 731 & $7.6\left(13\right)$ & $8.8\left(13\right)$ & $5.6\left(-8\right)$ & $5.9\left(-8\right)$ & -- & $3\left(-8\right)$ & 2\\
C$_3$S & 104 & 521 & $4.5\left(13\right)$ & $1.7\left(13\right)$ & $3.4\left(-8\right)$ & $1.1\left(-8\right)$ & -- & $1.2\left(-8\right)$ & 2\\
c-C$_3$H$_2$ & 244 & 1011 & $7.6\left(13\right)$ & $1.9\left(14\right)$ & $9\left(-8\right)$ & $2.1\left(-7\right)$ & $6.8\left(-7\right)$ & $5.6\left(-8\right)$ & 3\\
CH$_3$CN & 128 & 538 & $8.4\left(13\right)$ & -- & $4.9\left(-8\right)$ & -- & -- & $2.6\left(-8\right)$ & 3\\
CH$_2$CN & 383 & 1009 & $1.2\left(13\right)$ & $2.1\left(13\right)$ & $1.7\left(-8\right)$ & $2.8\left(-8\right)$ & -- & $1.9\left(-8\right)$ & 5$^a$\\
H$_2$CS & 244 & 870 & $4.9\left(13\right)$ & $4.8\left(13\right)$ & $3.6\left(-8\right)$ & $3.3\left(-8\right)$ & -- & $5.4\left(-9\right)$ & 5$^a$\\
\enddata
\tablecomments{Second and third columns denote the range of radii where the molecule is detected above 3$\sigma$, and also the distances over which abundances and column densities are averaged. Values are written with the exponent in parentheses. For each molecule observed toward V Hya, two column densities and abundances (relative to \ce{H2}) are listed, the first (subscript 1) was calculated assuming LTE and the cool temperature model, similar to the kinetic temperatures derived by \citet{Sahai2022_VHyaDUDE}; the other (subscript 2) uses the warm model obtained from the highest rotational temperatures found in our excitation analysis in Section \ref{section:methyl_pops}. For species with available collisional rates, non-LTE calculated abundance using RADEX is provided. Given the uncertainties in line flux, and the spread of abundances obtained over a wide range of temperature assumptions, the estimated error for these single-transition calculations is a factor of five. The last two columns list the observed abundance toward IRC+10216 and the reference for that value. $^a$Published column density converted to fractional abundance assuming source size matches that of \ce{CH3CN} [1] \citet{Cernicharo2010_IRC_HIFI_SiC2}, [2] \citet{Agundez2009_PhDT}, [3] \citet{Gong2015_1.3cm_IRC} [4] \citet{Tuo2024_IRC_survey} [5] \citet{Agundez2008_newIRCmols}}
\label{table:av_abunds}

\end{deluxetable*}

\begin{figure*}[t!]
    \centering
    \includegraphics[width=0.9\linewidth]{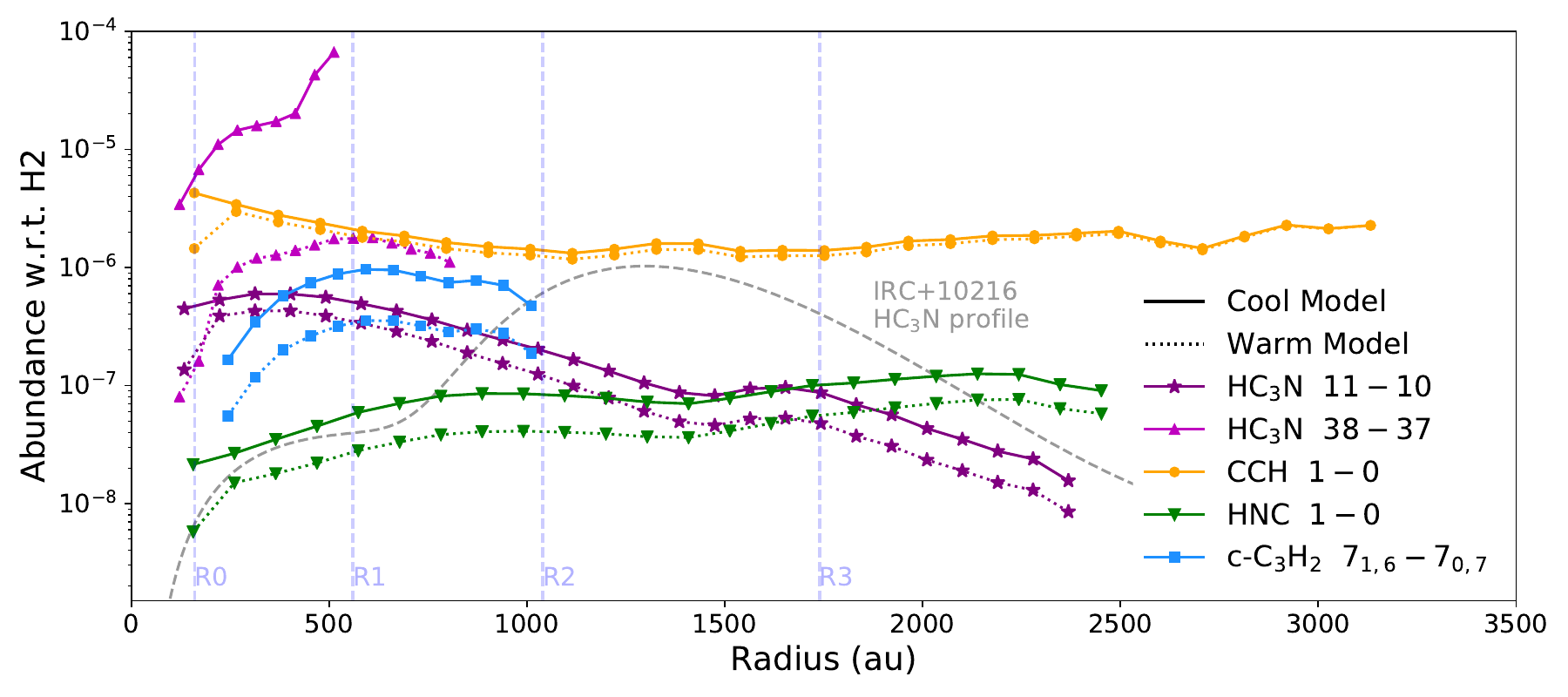}
    \caption{Non-LTE-simulated abundances of molecules in V Hya using RADEX, assuming purely collisional excitation (no vibrational pumping). Again, models fit assuming the cool temperature profile are shown with solid lines, while those fit using the warm temperature model are shown with dotted lines. Radial spacing of points is one-quarter of the beam size, and abundances are only calculated where the disk-averaged line flux is larger than 3 times the rms of the image. Since the uncertainty could not be carried through the forward-modeling process, it is assumed to be $30\%$, matching the typical error in radial integrated line flux.}
    \label{fig:NLTE_abunds}
\end{figure*}

The $J=38-37$ transition of \ce{HC3N} was omitted from the above LTE analysis, as \citet{Siebert2022_IRC_HC3N} found that both IR-excitation and subthermal excitation are important to consider for the excitation of this line. While IR-pumping is difficult to model in this source, we can investigate collisional non-LTE effects in a line-by-line manner using the radiative transfer code RADEX \citep{VanDerTak2007_RADEX}. For each radially extracted spectrum, we fit the integrated line flux to obtain the column density, then perform the same analysis on the \ce{^{13}CO} to convert this to an abundance. The kinetic temperature profiles used in the RADEX calculation are once again those shown in Fig.\ \ref{fig:pop_diags}, including the cool model derived in \citet{Sahai2022_VHyaDUDE} as well as the warm model we find from \ce{CH3CN} population analysis. The gas density profile for V Hya is also adopted from \citet{Sahai2022_VHyaDUDE}.

For the collisional excitation rates, we use the available molecular data from the Leiden Atomic and Molecular Database \citep[LAMDA;][]{vanderTak2020_lamda}. In the case of \ce{HC3N}, the collisional rates are only calculated up to $J=37$, so we extrapolate these up to $J=45$ using radiative transitions rates from CDMS \citep{2005JMoSt.742..215M}.

The results of this non-LTE analysis are shown in Fig.\ \ref{fig:NLTE_abunds}, and the ensuing average abundances are included in Table \ref{table:av_abunds}. The steep increasing abundances of the obtained from the $J=38-37$ transition of \ce{HC3N}, and the extremely high abundances ($>10^{-5}$) obtained when assuming a cool temperature model can be attributed to the high sensitivity of this line to gas density and temperature. Because the addition of vibrational pumping would increase the accessibility of higher rotational states, most of the $J=38-37$ abundances in Fig.\ \ref{fig:NLTE_abunds} can be considered upper limits. In contrast, when the $J=11-10$ transition is used to calculate the abundance it appears much closer to the profile found in the previous section, and the total average abundance is within 20\% of the cool-temperature model ($f_{\mathrm{LTE}1}$), meaning LTE is a valid approximation for these lower $J$ states. While the shapes of the $J=38-37$ and $J=11-10$ non-LTE profiles are quite different, the abundances within R0 (where collisions are expected to be the dominant excitation mechanism for \textit{both} transitions) are consistent under the warm temperature model, meaning ${\sim}3-5\times10^{-7}$ is a good estimate for the abundance of \ce{HC3N} in these central regions of the DUDE. Outside of this region, radiative effects must be considered to model the $J=38-37$ line properly.

Similar to the low-energy line of \ce{HC3N}, the calculated abundance of \ce{CCH} shows only a marginal change when RADEX is used. This is expected, as the transition rate of this line is quite low ($A_{ij}{\sim}10^{-6}$), so it is likely thermally populated throughout most of the DUDE. The overall shape of the abundance is also preserved in our RADEX calculations, and we continue to see that \ce{CCH} is abundant even at very small radii in V Hya.

The derived abundance profiles of \ce{HNC} and c-\ce{C3H2} show a clear shift to larger radii when LTE is not assumed. This can again be attributed to the reduced effectiveness of collisions at large radii (and consequently lower densities). For c-\ce{C3H2}, the large jump in its peak abundance (from ${\sim}10^{-7}$ to $10^{-6}$) in the non-LTE model indicates that the excitation temperature is much lower than the kinetic temperature. This case is likely similar to \ce{HC3N} $J=38-37$, where the transition rate is so high that collisions alone cannot populate the rotational levels, and an additional mechanism like IR-pumping is needed. For lines like these, the high temperature LTE approximation in the previous section is more accurate. In contrast, for the low-lying $J$-states sampled by Band 3 transitions (\ce{HNC}, \ce{CCH}, and \ce{HC3N} $J=11-10$), the non-LTE collisional model shown in Fig.\ \ref{fig:NLTE_abunds} is a good representation of the abundance distribution in V Hya.

\subsubsection{Chemistry in the Circumbinary Disk}
The observed molecular inventory of V Hya is not unlike other carbon-rich AGB envelopes, and the estimated abundances indicate that the established chemical processes for these environments are not drastically different in this unique morphology. At the same time however, the relatively compact distributions of daughter molecules in the flared disk indicate that photochemical processes are affected in this unique morphology. To maintain the observed abundances of \ce{CCH} near the center of the outflow, there must be a sufficient amount of dissociating radiation (typically UV) present in these regions. We propose three potential scenarios for the source of this excess radiation in the inner disk of V Hya. The first is an external origin. In a spherical AGB outflow, interstellar photons near the extended stellar atmosphere are attenuated by the entire envelope, resulting in a large degree of extinction and a slow build-up of photochemical products with radius. In the environment of V Hya though, the geometry of the DUDE also allows for ambient interstellar photons to penetrate from less obscured directions above and below the disk, which could reduce the $A_V$ at small radii significantly. This would also imply that there could be a vertical stratification of chemical species in the DUDE, similar to what is seen in protoplanetary systems \citep{Oberg2023_PPDchem_review}. 

Another explanation is that the chemistry-driving radiation is produced internally, either from a higher temperature companion, or from the evolved star itself if it is transitioning to the post-AGB phase and thus exposing a hot central star. The latter is the case for the C-rich protoplanetary nebula CRL 618, which harbors a central photon-dominated region (PDR) and initiates a fast-evolving photochemistry \citep{Woods2003_CRL618model}. V Hya shares kinematic similarities with CRL 618 with the presence of periodic high-velocity, bulleted-ejections; however, CRL 618 is B0-type and is clearly ionizing the inside of its envelope \citep{Fong2001_evolvedstarPDRs}, whereas V Hya is much more red \citep[C6;][]{Samus2017_variablecatalog} and shows no clear evidence of having left the AGB. 

V Hya is spectrally unique among carbon stars however, as it is observed to have a substantial time-variable near-UV excess in its spectral energy distribution. This could partially be attributed to an A0-type secondary star \citep{Planquart2024_VHya_comp_monitoring}, or from an accretion disk surrounding the companion \citep{Sahai2008_GALEX_UV_AGB,Montez2017_AGB_GALEX}. The latter has also been inferred as the origin of the high-velocity bipolar ejections seen in this source \citep{Hirano2004_VHya,Sahai2009_VHya,Scibelli2019_VHya_bulletModeling,Sahai2022_VHyaDUDE,Planquart2024_VHya_comp_monitoring}. Such an accretion disk or companion would produce internal source of chemistry-driving radiation that wculd impact the distributions of molecules presented in this work.

The hypothesis of chemistry induced by stellar companion photons was recently found to be important in IRC+10216 \citep{Siebert2022_IRC_HC3N}, the S-type AGB W Aql \citep{Danilovich2024_WAql_binarychem}, and previously by \citet{Vlemmings2013_RScl_COratios} and \citet{Saberi2018_Mira_CI} for the C- and M-type stars R Scl and omi Cet. It is well-known that V Hya has a companion with an orbital period of 17\,yr, a large eccentricity, and a proposed spectral type of A0 \citep{Knapp1999_VHya_evo,Planquart2024_VHya_comp_monitoring}. W Aql is also part of a very eccentric binary, and \citet{Danilovich2024_WAql_binarychem} found enhancements of \ce{SiN} in the location where periastron occurred with an F-type companion. While we do find evidence of some asymmetric abundance distributions in V Hya (i.e.\ \ce{C4H}), none are as stark as those seen in W Aql. We also note that in general, most daughter species in V Hya do not show any anisotropies in their emission beyond the expected projection effects of an inclined disk.

Without a detailed 3-dimensional radiative and chemical model of this object, it is unclear which of the above scenarios are the most important factors in the observed distributions of gas phase species toward V Hya. Nevertheless, the observational results provided here provide the initial chemical constraints which suggest that these dynamical effects do in fact play a role in shaping a unique molecular environment in this source. We therefore stress that these evolutionary factors should therefore be considered in future theoretical and observational efforts to understand chemistry in both V Hya and similarly complex AGB-related environments.

\section{Conclusions}

The environment of V Hya is a prime location to examine C-rich AGB chemistry in a complex, binary-induced circumstellar morphology. To this end, we have presented a spatio-chemical analysis of this object using available ALMA observations with the goal of mapping the gas phase molecular inventory in its expanding disk. We report rotational lines from 20 unique molecules and isotopologues, including parent molecules, polyyne carbon chains, and more saturated species like c-\ce{C3H2}, \ce{CH3CN}, and \ce{H2CS}. With a physical model of the DUDE in hand, we compared the spatial distributions of detected transitions of these molecules, and derived excitation conditions and initial estimates on abundances throughout this source. The main results of these analyses are summarized as follows:

\begin{enumerate}
    \item Emission from all detected molecules is primarily constrained within the first density-enhanced ring of the DUDE (R1; 560\,au). Low-energy transitions of \ce{HNC}, \ce{CCH}, and \ce{HC3N} can extend farther than this, to a maximum of ${\sim}$2000\,au. 
    \item Most lines from daughter species display single-peaked, ring-shaped spatial distributions. The radial positions of the intensity maxima vary among detected transitions. Parent species, and curiously \ce{CH3CN} exhibit centrally-peaked distributions.  
    \item The ``R0" ring identified from \ce{HC3N} emission by \citet{Sahai2022_VHyaDUDE} at 160\,au is detected in some other transitions like those of \ce{C4H}. However, the non-ubiquity of this ring among chemical species, as well as its absence from the \ce{^{13}CO} maps suggests that this is not a physical sub-structure in the disk, but a location that is favorable for the excitation of carbon chains. 
    \item Some molecules show peculiar asymmetries in their emission. This is most commonly seen in the many detected lines of \ce{C4H}, which all appear brighter in southern regions of the DUDE at a deprojected distance of ${\sim}200$\,au from V Hya. These anisotropies are not apparent in \ce{^{13}CO} nor other molecular lines tracing this region (\ce{HC3N}, \ce{CH3CN}), suggesting they reflect a chemical asymmetry in the disk.
    \item Using the known velocity-structure of the outflow, image cubes of V Hya can be effectively be shifted and stacked with a common line center to greatly boost the significance of rotational line detections. Using these methods, we confirm the detection of \ce{CH2CN} in V Hya, and separate emission from $l$-\ce{C3H} from a blended transition of \ce{CH3CN} (Appendix \ref{app:gofish_more}). This type of spectral cube manipulation could also be useful for other sources with well-constrained, 2D-projected kinematics. 
    \item A spatial excitation analysis of species with multiple observed lines yield a wide range of $T_{\mathrm{rot}}$ values. Species like \ce{SiC2} are found in agreement with the kinetic temperature model of the disk found in \citet{Sahai2022_VHyaDUDE}, while other molecules (\ce{CH3CN}, \ce{C4H}) have higher rotational temperatures by up to a factor of three. At the same time, the change in excitation temperature with radius is found to be consistent between species. Using these population fits, we place bounds on the excitation temperatures of molecules in this source, and use these new temperature models to place conservative estimates on the physical conditions of species that are detected with fewer lines.
    \item The single-line estimated average abundances of carbon chain species in the disk are consistent (to within a factor of ${\sim} 5$) with C-rich AGB envelopes of similar mass-loss rate to V Hya.
    \item The spatial distributions of molecular abundances are more compact surrounding V Hya than in comparable carbon star envelopes. All carbon chains reach their maximum abundance within a radius of 500\,au. Some daughter species (e.g.\ \ce{CCH}, \ce{C4H}, and \ce{CH3CN}) have strictly decreasing abundances with radius (save for a global drop inside 200\,au) and thus show little evidence of their expected shell-like abundance distributions. In both LTE and non-LTE estimates, \ce{CCH} must have a fractional abundance w.r.t.\ \ce{H2} larger than $10^{-6}$ in the inner ${\sim}150$\,au of the DUDE to reproduce the observed lines. For comparison, in IRC+10216 this abundance is only reached at radii larger than 1000\,au \citep{Agundez2017_IRC_carbonchains}. This result is quite different from the traditional model of AGB chemistry, which predicts first a steady increase in the abundance of photochemistry products with increasing radius, followed by their breakdown at larger radii.
\end{enumerate}

These results highlight both similarities and differences between this unique source and the more spherical morphologies of other well-studied carbon stars. It is presently unclear what chemical mechanisms are responsible for the close-in production of carbon chain species in the expanding disk of V Hya, as well as the drop in all molecular abundances including CO below 200\,au. To resolve this, further analysis of this source (and similarly-perturbed AGBs) will be key, including large bandwidth surveys covering many transitions of the species investigated in this work, and 3D chemical modeling efforts for these wind structures. In particular, we stress the need to characterize small photodissociation products and carbon-bearing radicals (e.g.\ \ce{CCH}, \ce{C4H}, \ce{CN}) in the innermost regions of stellar winds, as these are crucial to initiating the formation of larger species, and also appear to have some of the most peculiar behavior in V Hya. 

Equatorial circumstellar density structures have been increasingly common in the study of AGB stars \citep{Decin2020_atomium}, and are well-established in more evolved systems \citep{Sahai2007_PPNsurvey_class,Olofsson2019_HD101584}. As observational capabilities grow, chemical case studies of rapidly evolving AGB and RSG sources will be key in understanding the full scope of molecular processes in evolved stars, and how they are coupled with complex mass-loss mechanisms. Pursuing this broad characterization of circumstellar chemistry will in turn allow us to place important constraints on the recycling of stellar material, and the budget of complex species which can then be inherited by the ISM.

\section{Acknowledgements}
We would first like to thank the anonymous referee for their thorough and constructive review of this report. This paper makes use of the following ALMA data: ADS/JAO.ALMA\#2019.1.00507.S, ADS/JAO.ALMA \#2015.1.01271.S, ADS/JAO.ALMA\#2018.1.01113.S. ALMA is a partnership of ESO (representing its member states), NSF (USA) and NINS (Japan), together with NRC (Canada), MOST and ASIAA (Taiwan), and KASI (Republic of Korea), in cooperation with the Republic of Chile. The Joint ALMA Observatory is operated by ESO, AUI/NRAO and NAOJ. The National Radio Astronomy Observatory is a facility of the National Science Foundation operated under cooperative agreement by Associated Universities, Inc. Support for this work was provided to M.\,A.\,S.\, by the NSF through the Grote Reber Fellowship Program administered by Associated Universities, Inc./National Radio Astronomy Observatory. M.\,A.\,S.\, also received support under contract at Chalmers University of Technology during the review period of this work. M.\,A.\,S.\, finally acknowledges additional support from the Virginia Space Grant Consortium. R.S.’s contribution was carried out at the Jet Propulsion Laboratory, California Institute of Technology, under a contract with NASA.

\bibliography{refs}
\bibliographystyle{aasjournal}

\appendix

\renewcommand{\thefigure}{A\arabic{figure}}
\renewcommand{\thetable}{A\arabic{table}}
\renewcommand{\theequation}{A\arabic{equation}}
\setcounter{figure}{0}
\setcounter{table}{0}
\setcounter{equation}{0}

\section{Further applications of line stacking}
\label{app:gofish_more}

\begin{figure*}[h!]
    \centering
    \includegraphics[width=0.95\linewidth]{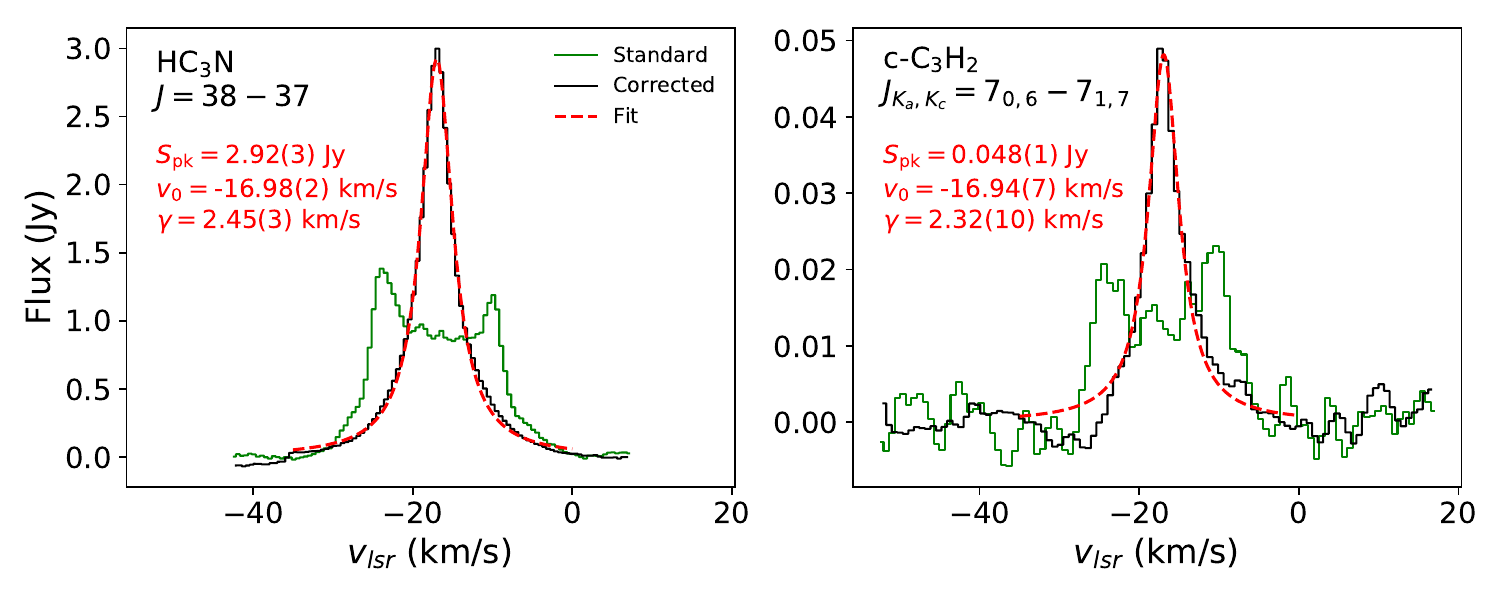}
    \caption{Original (solid green) and velocity-stacked (solid black), spatially-integrated line profiles for \ce{HC3N} and c-\ce{C3H2} toward V Hya. Dashed red line profile shows a Lorentzian fit to the corrected spectrum. Best fit values for the parameters in Eq.\ \ref{eq:lorentz} are labeled, with last-digit errors shown in parentheses. All spectra were obtained over a deprojected radius range from 0\arcsec--1.5\arcsec (600\,au), and therefore neglect emission outside of R1.}
    \label{fig:gofish_fits}
\end{figure*}

\begin{figure*}[t!]
    \centering
    \includegraphics[width=0.95\linewidth]{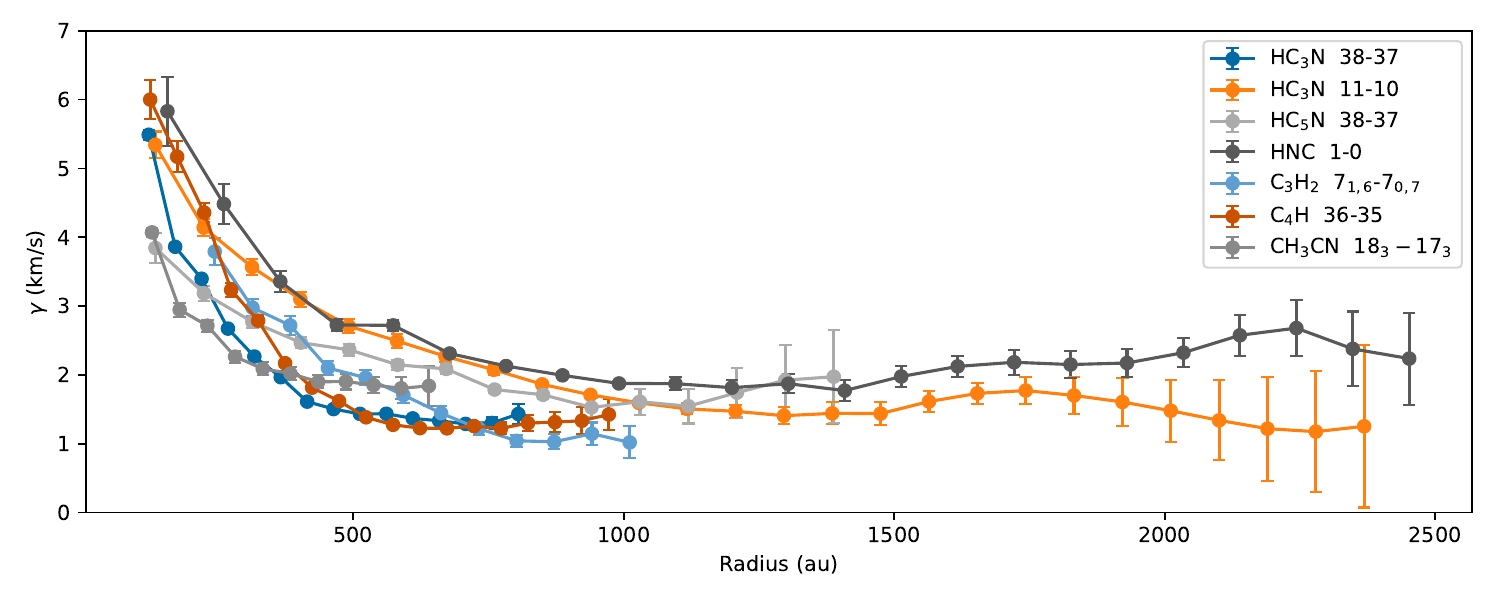}
    \caption{Measured line width ($\gamma$ in Equation \ref{eq:lorentz}) of stacked molecular line spectra extracted at observed radial distances from V Hya. The observed and simulated spectra corresponding to these fits can be found in Appendix \ref{app:all_fits}. Radial bin widths are equal to one-quarter of the ALMA beam for a given line.}
    \label{fig:gamma_radial}
\end{figure*}

The spectral stacking method first developed for protostellar sources, and described in Section \ref{section:disc}, has proven a useful tool for examining weak line signatures and from this evolved star. Here, we explore the results of this method in detail, and investigate additional use cases for it in the context of V Hya, as well as objects with similar morphology. Fig.\ \ref{fig:gofish_fits} shows the original and expansion velocity-corrected spectra of two example transitions (\ce{HC3N} and c-\ce{C3H2}). We find that the line shapes of the corrected spectra are not Gaussian, but are instead well fit by a Lorentzian profile:

\begin{equation}
\label{eq:lorentz}
    S(v)=\frac{S_{\mathrm{pk}}}{1+\left(\frac{v-v_0}{\gamma}\right)^2}
\end{equation}

This line shape is remarkably different from the Gaussian profiles retrieved in the protoplanetary disk applications of this method \citep{Teague2018_TWHya_CS_stack}, and is typically characteristic of pressure broadening; however, we do not expect that this is what is represented by the corrected line width in Fig.\ \ref{fig:gofish_fits}. Instead, we recall that the DUDE has a 3D flared structure and is not the infinitely thin disk assumed by the stacking method. Because the DUDE is rather thick, with an opening angle of ${\sim}39^{\circ}$, this effect likely dominates any other sources of line broadening, so the reduction in line-width is not as dramatic as it is in the case of protoplanetary disks \citep[e.g.\ Fig.\ 3 in][]{Yen2016_stackingdisk}. The additional flux in the line wings could also be contributed by the base of the high-velocity outflow in V Hya, which is not accounted for in the velocity model used for stacking, and therefore could result in a profile that appears Lorentzian in shape.

Figure \ref{fig:gamma_radial} shows the relationship between best-fit line width and radius for several observed molecules. All species show a characteristic decreasing trend with radius that is steeper at close distances to V Hya. This can be attributed to the fact that as one moves inward with radius, the prescription of a purely radially expanding DUDE with constant velocity begins to break down. At short inner radii, flux can also be contributed from the base of the high-velocity outflow and wind acceleration regions, which will not be shifted to the systemic velocity by this method, and thus will broaden the stacked profile. This is what produces the universal trend of increasing line width at short radii, and also could cause the more irregular line shapes observed in these regions.

We also find a high degree of variance among the detected transitions in Fig.\ \ref{fig:gamma_radial}, as $\gamma$ can vary by up to a factor of three at a given radius depending on the transition observed. As discussed above, with the expansion velocity of the mid-plane removed via spectral stacking, the main broadening mechanism for these lines is the velocity discrepancy from radial motion of material above and below the mid-plane along a particular line of sight. Therefore, the spread in line widths in Fig.\ \ref{fig:gamma_radial} supports the conclusion that these transitions trace variable heights in the DUDE. This conclusion is supported by the fact that the lines with the largest radial extent (\ce{HNC}, \ce{HC3N} $J=11-10$) also have the largest observed line widths (and therefore vertical extent).

We also note that in Fig.\ \ref{fig:gofish_fits}, the best-fit lsr velocities are well-constrained at $-16.96$--$.98$\,\kms. This is also consistent with the observed peaks of \ce{C4H} and \ce{CH2CN} in Fig.\ \ref{fig:gofish_stack}. Since the shifting and stacking algorithm only accounts for the motion of gas relative to V Hya, this value can be interpreted as a more accurate value for the system velocity (elsewhere quoted in this work as $-17.4$\,\kms).

\begin{figure*}[t!]
    \centering
    \includegraphics[width=0.95\linewidth,trim=0.3cm 0.7cm 0.5cm 1.2cm,clip]{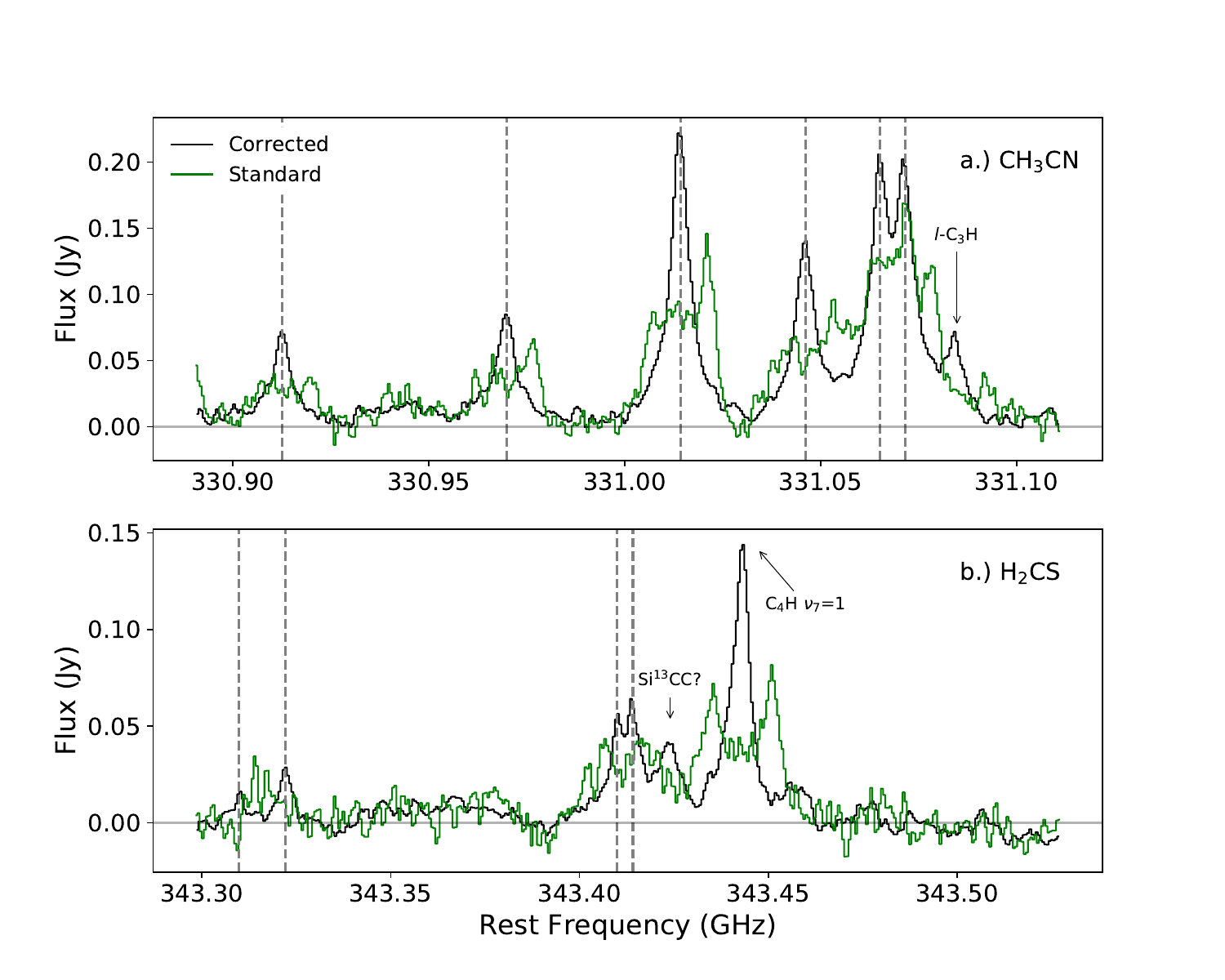}
    \caption{Application of line shifting and stacking procedure on larger bandwidths covering groups of \ce{CH3CN} (top) and \ce{H2CS} (bottom) transitions. For these molecules, we indicate the catalog rest frequencies \citep[CDMS; ][]{2005JMoSt.742..215M} with vertical dashed lines. From left to right, these represent the $K=5,4,3,...,0$ components of the $J=18-17$ transition of \ce{CH3CN}, and the $J_{K_a,K_c}=10_{4,7}-9_{4,6}$, $10_{2,9}-9_{2,8}$, $10_{3,8}-9_{3,7}$, $10_{3,7}-9_{3,6}$ lines of \ce{H2CS}. Additional blended species are labeled with arrows. Both the corrected and uncorrected spectra were converted to rest frequency using a modified systemic velocity of $v_{lsr}=-16.96$\,\kms, and were extracted from the central region of the DUDE (0\arcsec--1.5\arcsec).}
    \label{fig:gofish_blends}
\end{figure*}

Applying these position-dependent velocity corrections also is a useful way to investigate blended regions of emission lines. This is because the emission can be partially decoupled by moving spectra on either side of the DUDE away from each other in velocity space. The best examples of this are apparent in the \ce{CH3CN} and \ce{H2CS} groups of blended lines, shown both corrected and uncorrected in Fig.\ \ref{fig:gofish_blends}. For \ce{CH3CN}, we can clearly see the peaks of K=0, K=1, and K=2 in the stacked spectrum, which were previously not distinguished in the original spectrum due to the width and complexity of the original line profile. The positions of these peaks agree to within 1\% of the catalog rest frequency when shifted using the modified system velocity of $-16.98$\,\kms. Additionally, this method reveals the detection of the $J=21/2-19/2$ transition of $l$-\ce{C3H}, which was only marginally seen in the uncorrected spectrum. 

The \ce{H2CS} lines shown in Fig.\ \ref{fig:gofish_blends}b.\ are also obvious after stacking. The $10_{3,8}-9_{3,7}$, $10_{3,7}-9_{3,6}$ transitions of this molecule are the nearest components that we separate with this method,  only 3.6\,\kms (4.2\,MHz), less than 25\% the original double-peaked profile width. The stacked spectrum also improves the S/N of the much weaker $10_{4,7}-9_{4,6}$ and $10_{2,9}-9_{2,8}$ \ce{H2CS} lines, and reveals another feature that was previously buried among the \ce{H2CS} and \ce{C4H} $\nu_7=1$ transitions centered at 343\,424\,MHz. We attribute this to the $15_{9,7}-14_{9,6}$ transition of \ce{Si^{13}CC}. 

While stacking is incredibly useful in de-blending molecular signatures, care must be taken when choosing the bandwidth over which this is applied. Because the appropriate velocity shift of each pixel is dependent on the rest frequency of a given line, its accuracy decreases for lines at the edge of a large band. Another caveat is that this method is only applicable to objects where every line of sight in an image can be mapped to a single (central) lsr velocity, which is the case for an inclined, expanding disk as we have approximated V Hya. This means it is unfortunately not possible for a spherical AGB envelope, because the emerging emission profile from a particular line of sight in such a source is always centered on the systemic velocity, and thus there is no valid shift in frequency that can be used to align spectra at different positions in an image. The analysis is thus restricted to equatorial structures in sources with well-characterized expansion or rotation, and we recommend this approach for future interferometric spectral line analyses of AGB-related sources exhibiting such velocity gradients.

\section{Radial Spectral Fits for Abundance Calculation}
\label{app:all_fits}

In order to calculate molecular column densities and abundances, we perform an LTE fit to the radially extracted velocity-stacked spectra. Because the line shape is consistent among observed lines, spectra were simulated assuming the profile in Eq.\ \ref{eq:lorentz}. However, due to the variance in line width, when fitting the model, we leave $\gamma$ as a free parameter. Therefore, the resulting spectrum represents a minimization of $\chi^2$ in the parameter space of $N_T$ and $\gamma$. This fit was constrained to a 25\,\kms window centered on the system velocity, to ensure no nearby lines influence the minimized parameters.

\begin{figure*}[h!]
    \centering
    \includegraphics[width=\linewidth,trim=3.5cm 1.5cm 3.5cm 1cm,clip]{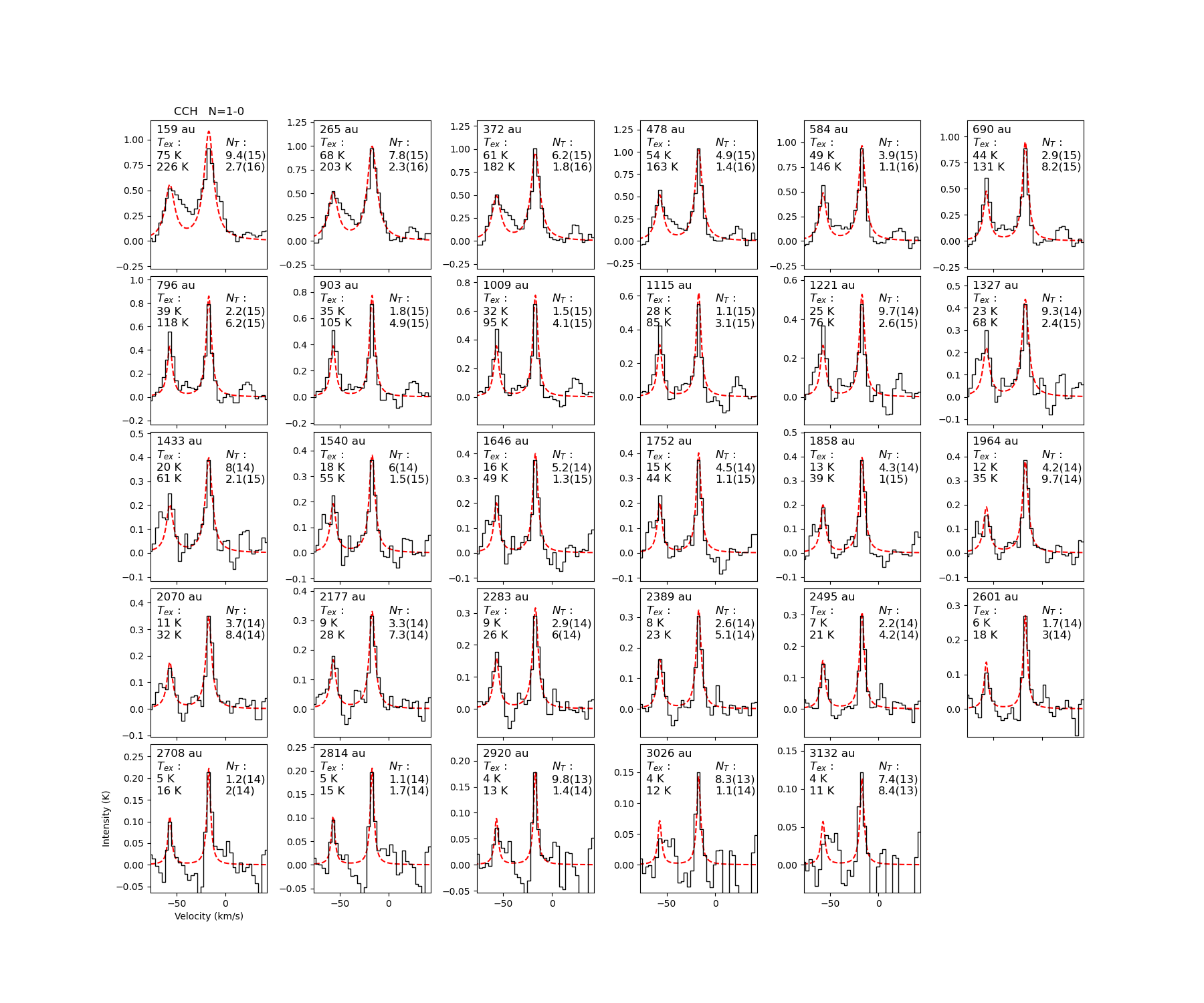}
    \caption{Fixed temperature LTE model (dashed red) of \ce{CCH} $N=1-0$ transitions at 87.317 and 87.329\,GHz compared with radial, velocity-stacked spectra observed toward V Hya (black). On each panel, column densities for the molecule are listed in units cm$^{-2}$, with the exponent in parentheses. The two column densities correspond to the best-fit LTE model when assuming the cool (top value) and warm (bottom) excitation temperature models of the DUDE. The temperatures of these models are shown on the left side of each spectrum.}
    \label{fig:fit_cch}
\end{figure*}

\begin{figure*}[h!]
    \centering
    \includegraphics[width=\linewidth,trim=3.5cm 1.5cm 3.5cm 1cm,clip]{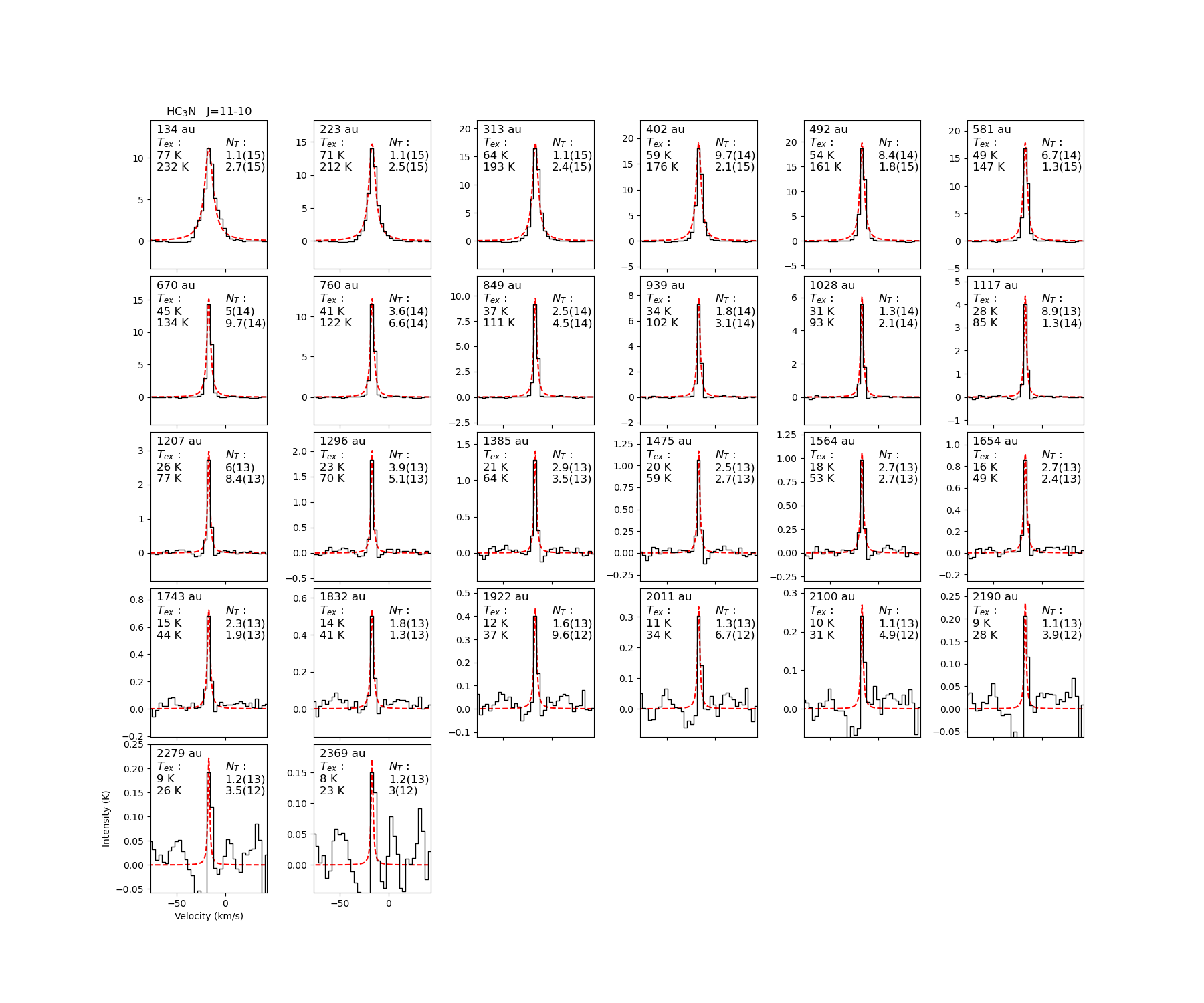}
    \caption{Same as Fig.\ \ref{fig:fit_cch} for the $J=11-10$ transition of \ce{HC3N} at 100.076\,GHz.}
    \label{fig:fit_hc3n}
\end{figure*}

\begin{figure*}[h!]
    \centering
    \includegraphics[width=\linewidth,trim=3.5cm 1.5cm 3.5cm 1cm,clip]{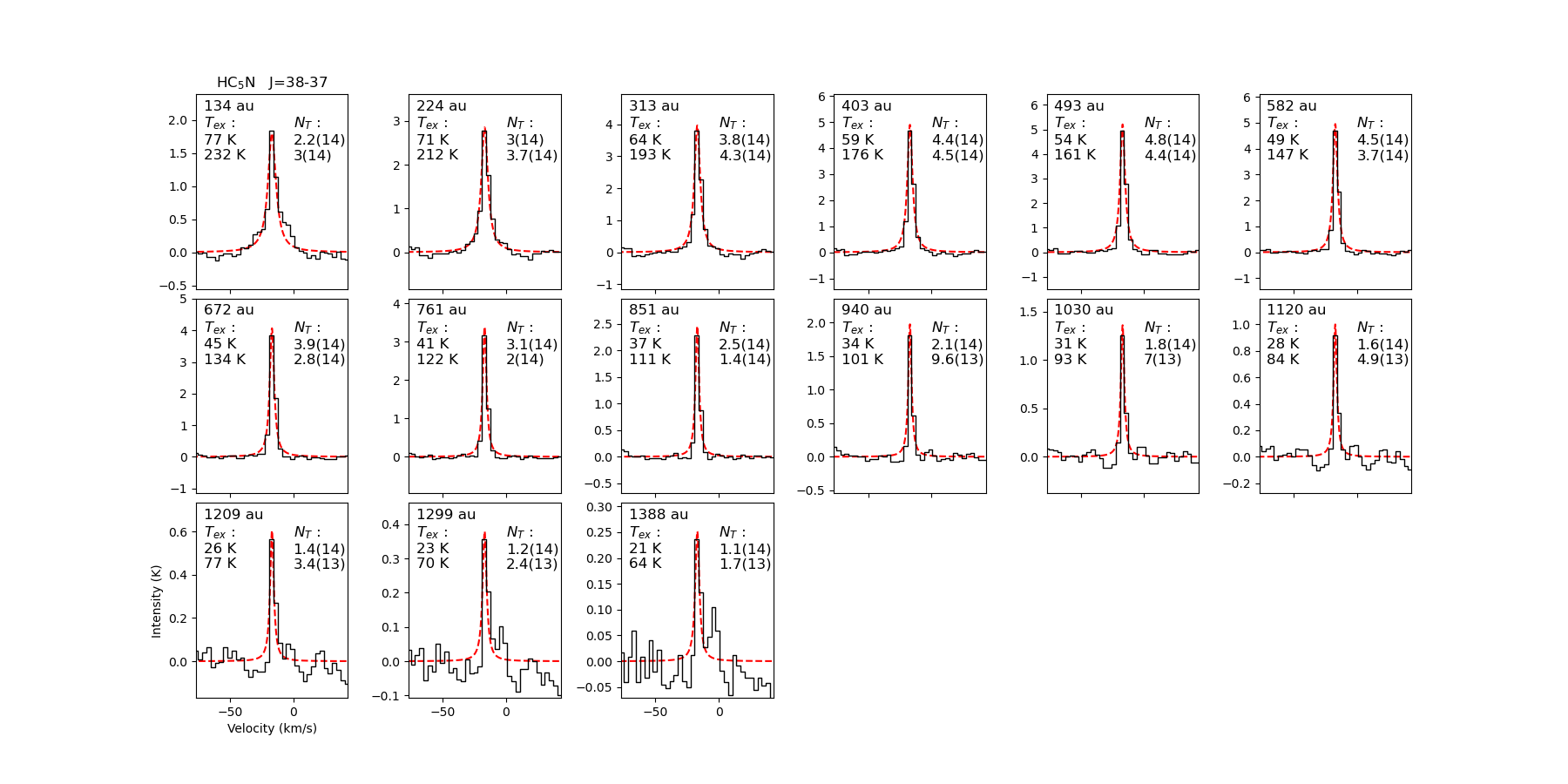}
    \caption{\ce{HC5N}: $J=38-37$ (101.175\,GHz)}
    \label{fig:fit_hc5n}
\end{figure*}

\begin{figure*}[h!]
    \centering
    \includegraphics[width=\linewidth,trim=3.5cm 1.5cm 3.5cm 1cm,clip]{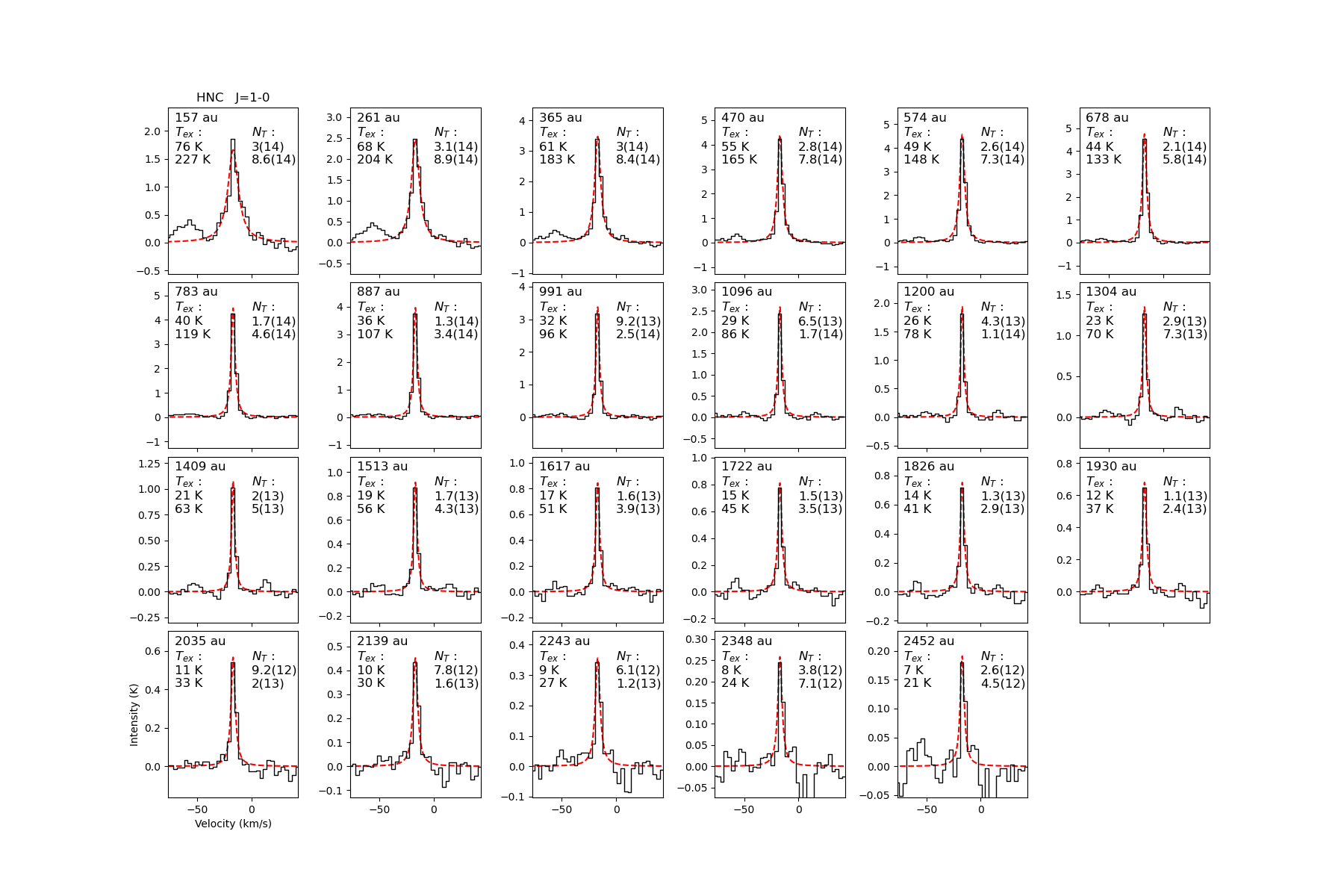}
    \caption{\ce{HNC}: $J=1-0$ (90.664\,GHz)}
    \label{fig:fit_hnc}
\end{figure*}

\begin{figure*}[h!]
    \centering
    \includegraphics[width=\linewidth,trim=3.5cm 1.0cm 3.5cm 0.5cm,clip]{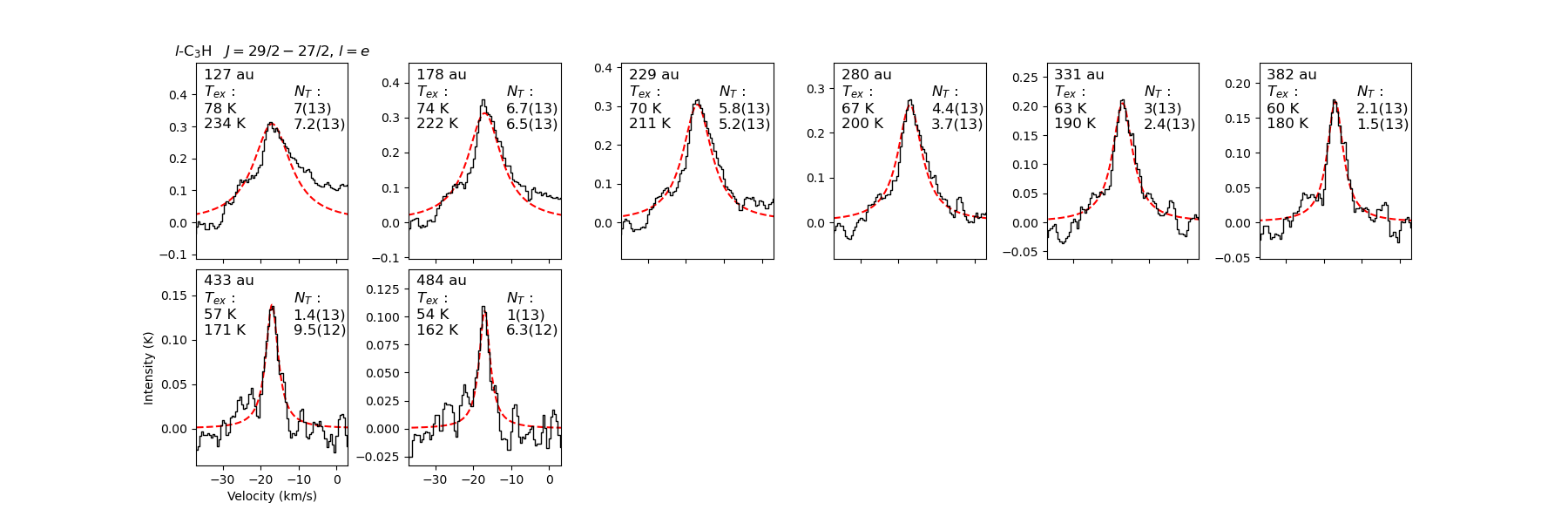}
    \caption{\ce{l-C3H}: $J=29/2-27/2$, $l=e$ (331.446\,GHz)}
    \label{fig:fit_c3h}
\end{figure*}

\begin{figure*}[h!]
    \centering
    \includegraphics[width=\linewidth,trim=3.5cm 0.5cm 3.5cm 1cm,clip]{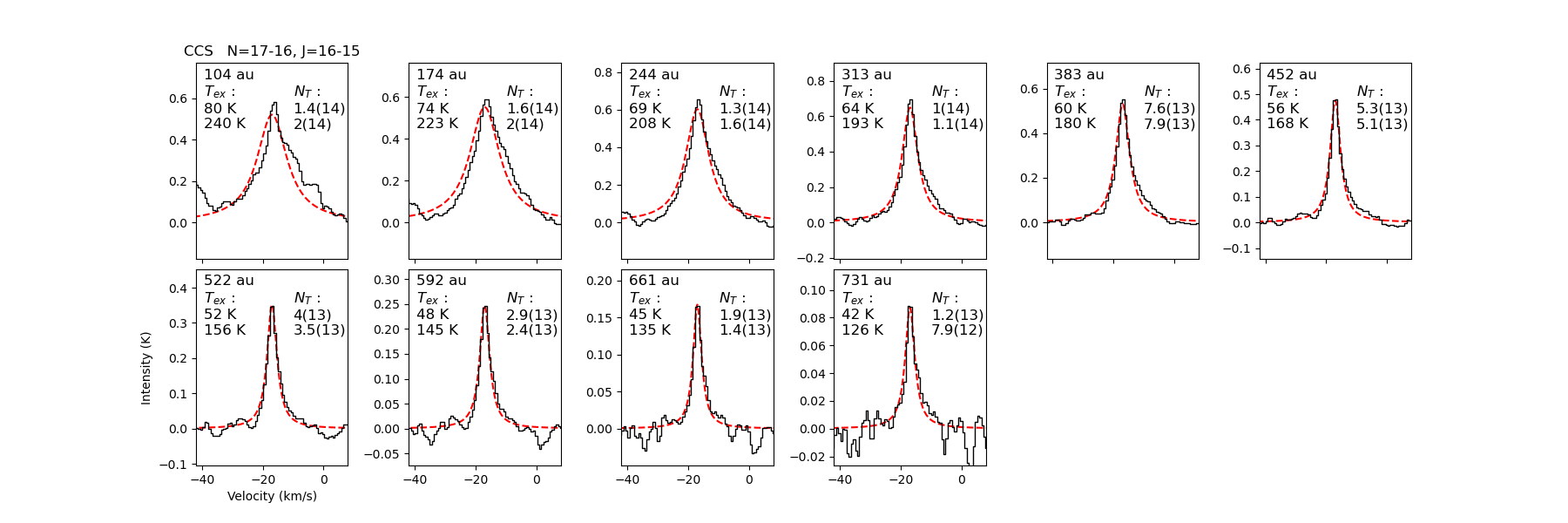}
    \caption{\ce{CCS}: $N=17-16$, $J=16-15$ (219.143\,GHz)}
    \label{fig:fit_ccs}
\end{figure*}

\begin{figure*}[h!]
    \centering
    \includegraphics[width=\linewidth,trim=3.5cm 0.5cm 3.5cm 1cm,clip]{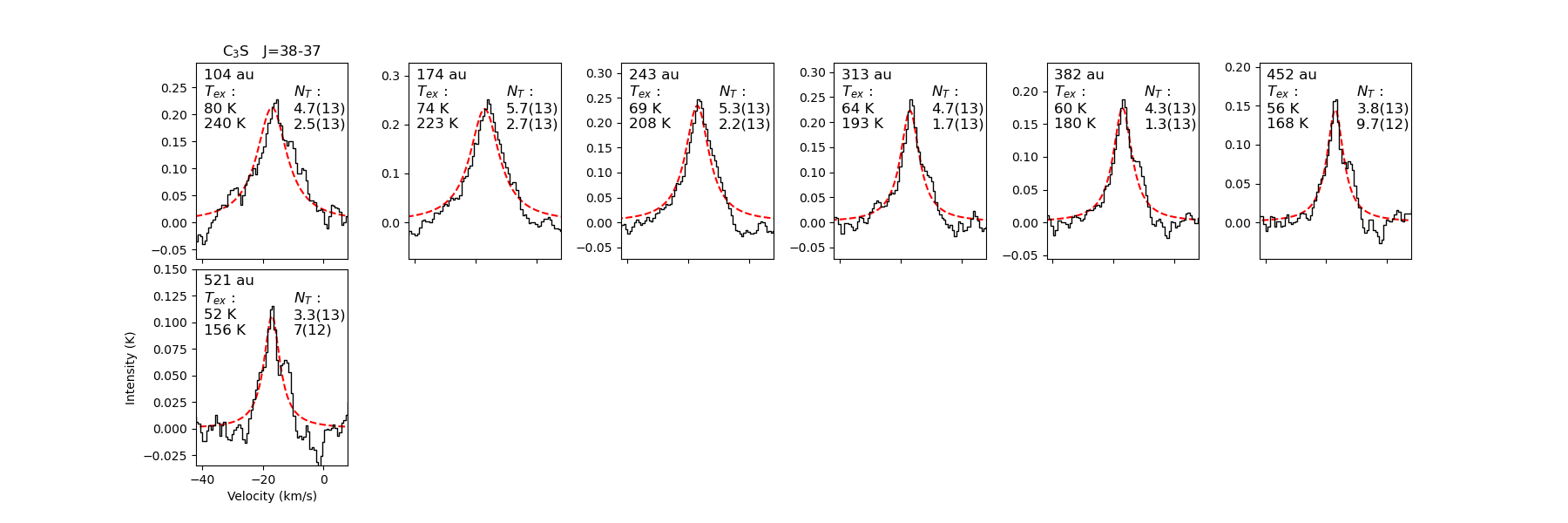}
    \caption{\ce{C3S}: $J=38-37$ (219.620\,GHz)}
    \label{fig:fit_c3s}
\end{figure*}

\begin{figure*}[h!]
    \centering
    \includegraphics[width=\linewidth,trim=3.5cm 0.5cm 3.5cm 1cm,clip]{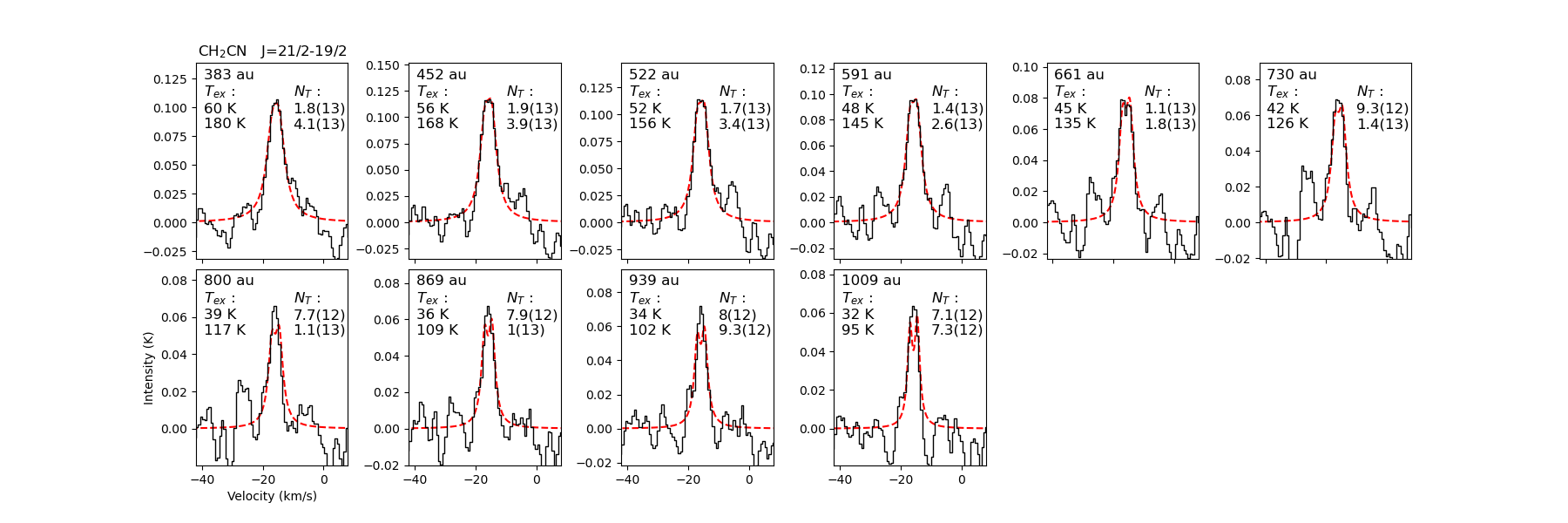}
    \caption{\ce{CH2CN}: $J=23/2-21/2$ (219.267\,GHz) and $J=21/2-19/2$ (219.269\,GHz)}
    \label{fig:fit_ch2cn}
\end{figure*}

\begin{figure*}[h!]
    \centering
    \includegraphics[width=\linewidth,trim=0cm 1.5cm 0cm 1cm,clip]{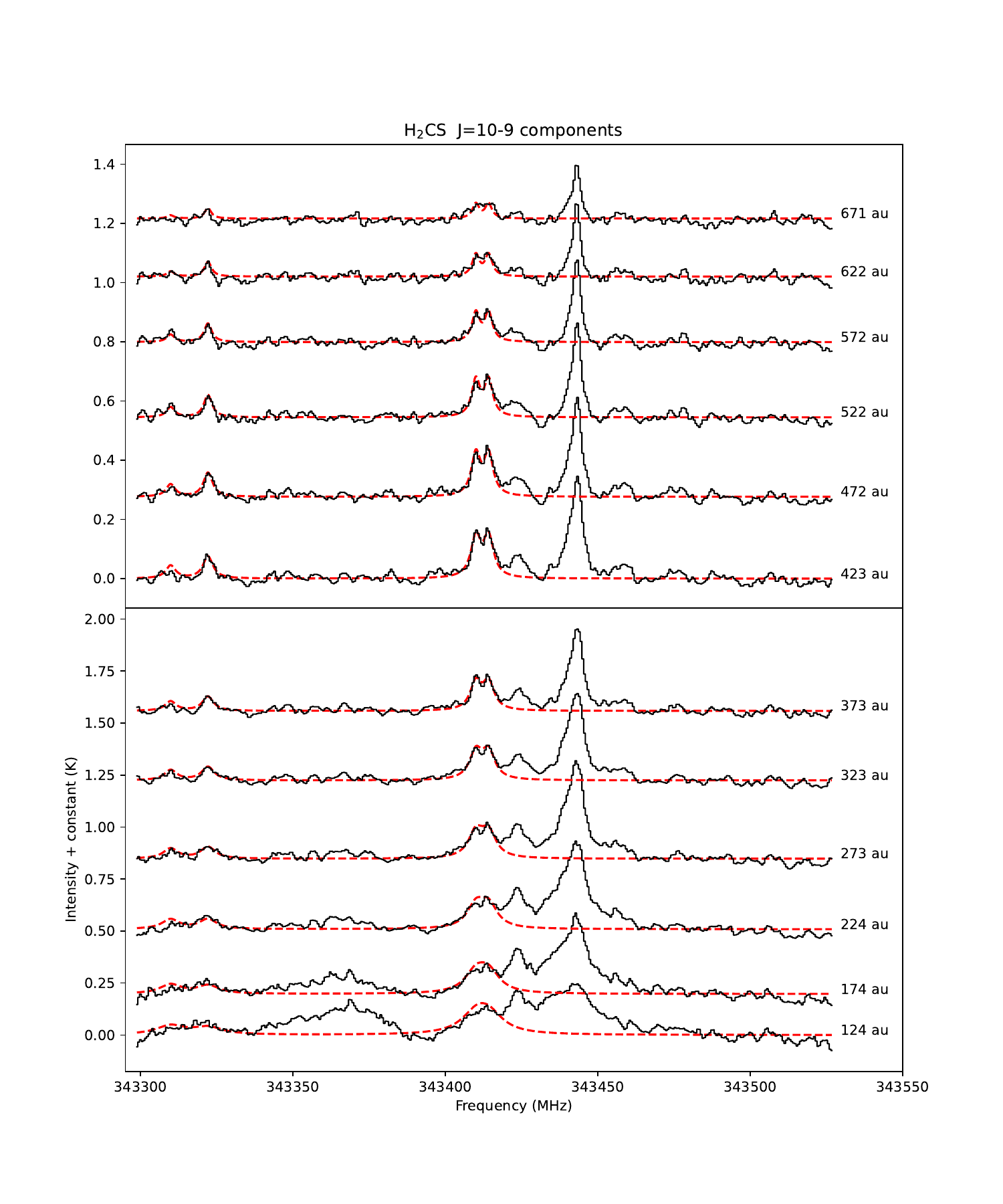}
    \caption{Multi-line spectral fit to the \ce{H2CS} $J=10-9$ group of transitions at 343.4\,GHz. The temperature is fixed to the cool temperature model for V Hya.}
    \label{fig:fit_h2cs}
\end{figure*}

Figures \ref{fig:fit_cch}--\ref{fig:fit_h2cs} show comparisons between the best-fit LTE-simulated and observed radial spectra for each studied transition (or group of transitions). Only one simulation is shown for each line (at each radius), as the model spectra assuming the cool and warm temperature models are identical.

One limitation of this method is that while the line shape is very predictable for large radii, this approximation appears to break down at short radii (e.g.\ Figures \ref{fig:fit_c3h} and \ref{fig:fit_ccs}). Additionally, because we use the $J=3-2$ line of \ce{^{13}CO} to convert column densities to fractional abundances, it is possible for there to be a mismatch in the vertical extent. Therefore, the listed abundances should be considered average abundances across disk heights where \ce{^{13}CO} emission is present at velocities $\pm10$\,\kms.

\begin{figure*}[h!]
    \centering
    \includegraphics[width=\linewidth,trim=3.5cm 0.5cm 3.5cm 1cm,clip]{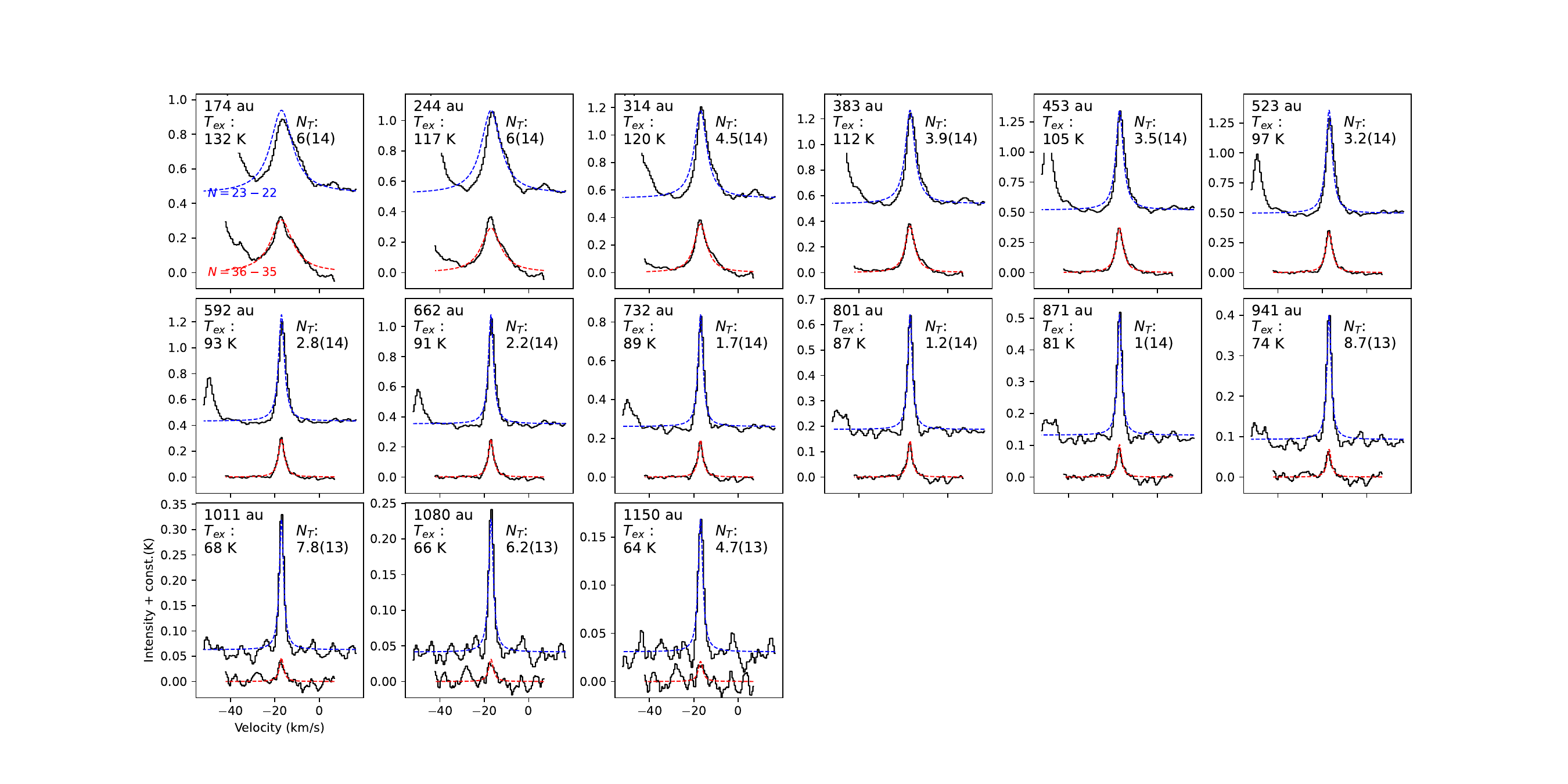}
    \caption{LTE simulation (dashed curves) of observed \ce{C4H} lines toward V Hya (solid) using the radial column densities and excitation temperatures from the two-point rotation diagram analysis presented in Section \ref{section:methyl_pops}. In each panel, the upper line is the $N=23-22$ ($J=47/2-45/2$) transition at 218.875\,GHz, and the lower is the $N=36-35$ ($J=73/2-71/2$) transition at 342.443\,GHz. The radii where spectra were extracted are shown at the top of each panel, along with the column density and temperature measured from the rotation diagram. The line width was set using the measured relationship between $\gamma$ and radius for velocity-stacked \ce{C4H} lines in this source.}
    \label{fig:fit_c4h}
\end{figure*}

\begin{figure*}[h!]
    \centering
    \includegraphics[width=\linewidth,trim=3.5cm 0.5cm 3.5cm 1cm,clip]{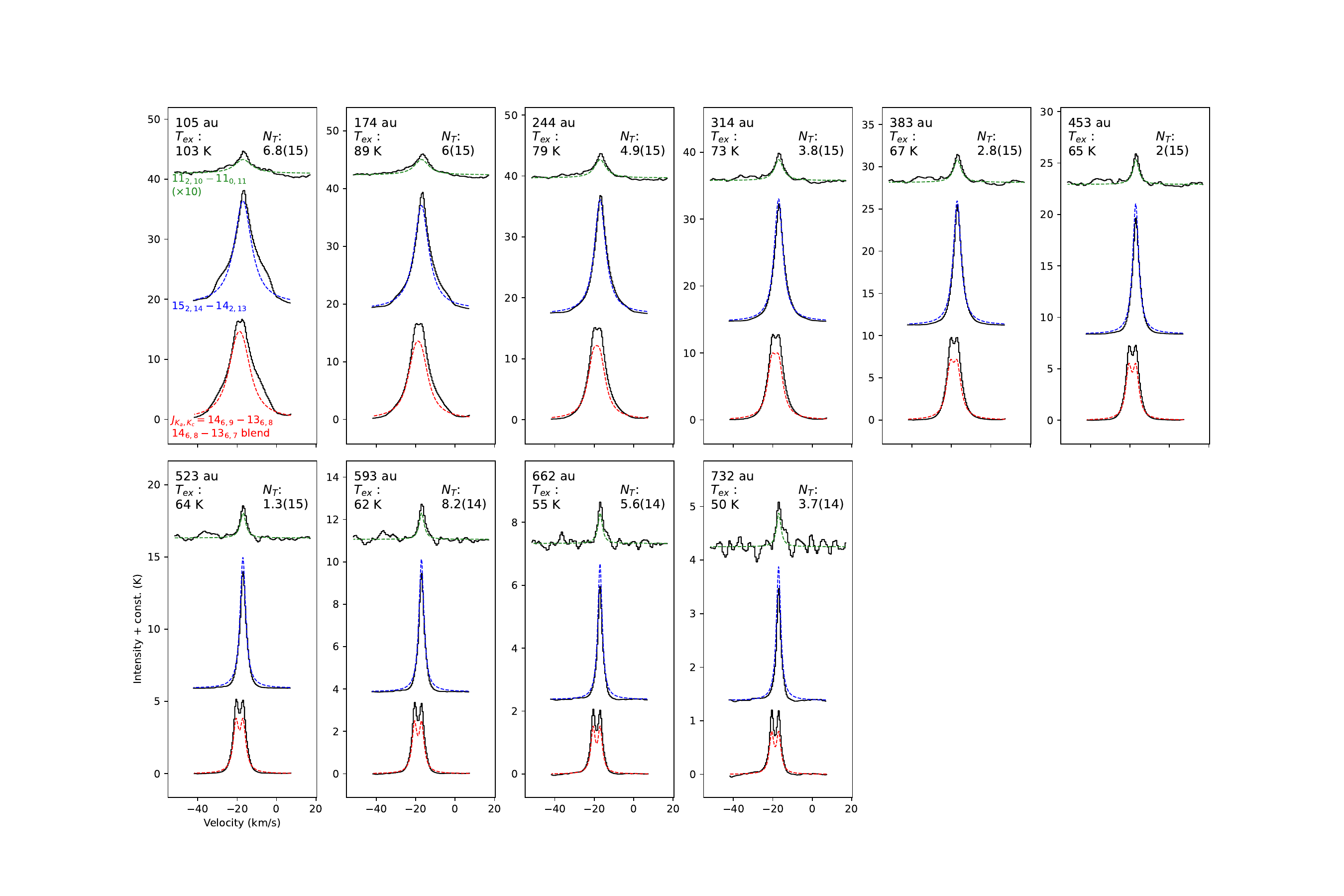}
    \caption{LTE simulation (dashed curves) of observed \ce{SiC2} lines toward V Hya (solid) using the $N_T$ and $T_{ex}$ from the rotation diagram analysis in Section \ref{section:methyl_pops}. From top to bottom, transitions are $J_{K_a,K_c}=11_{2,10}-11_{0,11}$ (218.787\,GHz), $15_{2,14}-14_{2,13}$ (342.805\,GHz), and a blend of $14_{6,9}-13_{6,8}$ and $14_{6,8}-13_{6,7}$ (330.870\,GHz, 330.874\,GHz). The radius, column density, and temperature measured from the rotation diagram are again showed at the top of each panel. The line width was set using the measured relationship between $\gamma$ and radius.}
    \label{fig:fit_sic2}
\end{figure*}

In Figures \ref{fig:fit_c4h} and \ref{fig:fit_sic2}, we show the LTE models corresponding to the rotation diagram fits for \ce{C4H} and \ce{SiC2} discussed in Section \ref{section:methyl_pops}.


\end{document}